\documentstyle[preprint,aps]{revtex}
\begin{document}
\tighten
\title{Jost Function for Coupled Partial Waves}
\author{
S. A. Rakityansky\footnote{Permanent address:
Joint Institute  for Nuclear Research, Dubna, 141980, Russia.}
and S. A. Sofianos}
\address{ Physics Department, University of South Africa,
 P.O.Box 392, Pretoria 0003, South Africa}
\date{\today}
\maketitle
\begin{abstract}
An exact method for direct calculation of the Jost functions
and Jost solutions for non-central potentials which couple 
partial waves of different angular momenta is presented. 
A combination of the variable--constant method
with the complex coordinate rotation is used to replace the matrix
Schr\"odinger equation by an equivalent system of linear first--order
differential equations. Solving these equations numerically,
the Jost functions can be obtained to any desired accuracy for all
complex momenta of physical interest, including the spectral points
corresponding to bound and resonant states. The effectiveness of the
method is demonstrated using the Reid soft--core and Moscow  
nucleon--nucleon potentials which involve  tensor forces. 
\end{abstract}
\vspace{.5cm}

\section{Introduction}

Almost any textbook on the scattering theory has a chapter devoted to the
Jost function, but none of them gives a practical recipe for its
calculation;  they provide instead equations expressing the Jost function in
terms of the wave function. However, to make use of them one must 
find the wave function first which means that the  problem is practically
solved and nothing more is needed. Thus  one usually gets a feeling
that the Jost function is a pure mathematical object,  elegant and
useful in formal theory, but impractical in computations. This is even 
more pronounced in  problems with non-central potentials which couple 
partial waves of different angular momenta. To the best of our knowledge, 
the classical book by Newton \cite{newton} is  the only one where  
the coupled partial wave Jost function, which in fact is a 
matrix--function, is considered and  the only calculation of the 
Jost matrix was done in Ref. \cite{sasakawa}. The need of 
such calculations  is of course indisputable since many potentials 
describing interactions between molecules, atoms, and nuclei 
are non-central.\\

It is a widely believed that the Jost function for partial waves
coupled by potentials  which are non-zero at the origin, is singular 
and therefore impractical. However, it is not the Jost function
but the Jost solution that diverges at $r=0$. Of course if one wants 
to obtain the Jost function via the Jost solution, in analogy
to un-coupled waves, then the problem of singularity is
inevitable and thus one must resort to alternative methods.\\

 In this work, we show that the Jost function can be calculated 
directly by simply solving certain first order coupled differential 
equations. These equations are based on the variable-phase 
approach\cite{calog} and their solution at any fixed value of the 
radial variable $r$, provides the Jost function and its complex 
conjugate counterpart, which correspond to the potential truncated at 
the point $r$.  Such  equations were proposed in the past for  
bound and scattering state calculations, that is for calculations 
in the upper  half of the complex momentum plane \cite{rakpup}.\\

Exploiting the idea of the complex rotation of the radial coordinate 
\cite{ccr}, $r\to r\exp(i \theta)$, the method was extended  to the
unphysical sheet so that the resonance state region was also
included~\cite{nuovocim}. This was possible because the 
coordinate rotation with a sufficiently large $\theta > 0$, makes 
the resonance state wave function quadratically integrable 
while the energies and widths of the bound and resonance states  
are not affected  since the Jost function and  the 
positions of its zeros do not
depend on $r$. The  effectiveness of the method was demonstrated in
Ref.~\cite{nth9607028} for potentials
with and without a Coulomb tail where high accuracy for very
narrow and very broad resonances was obtained. Its suitability in locating 
Regge poles in the complex angular momentum plane was also 
demonstrated.
In this paper we generalise the method to non-central potentials that
couple different partial waves and show that the complex rotation 
enables us to calculate the corresponding Jost matrix at all points 
of the complex momentum plane of physical interest. In a numerical 
example we  demonstrate that a high accuracy can also  be achieved.\\

The paper is organised as follows. In Sec. II and III our formalism  is
presented, while in Sec. IV  the method is applied to several examples and
the results obtained are discussed. Our conclusions are given in Sec. V.
Finally some mathematical details and proofs are given in the Appendix.
\section{Formalism}
Consider the system of two particles interacting via a non-central
potential $V({\vec r})$. Such potentials appear in many  physical
problems as for example in collisions of elementary and composite
particles with non-spherical molecules and  in problems involving
spin--dependent forces. \\

Since this type of  potentials are not rotationally invariant, the 
angular momentum $\ell$ associated with the interparticle coordinate 
${\vec r}$, is not conserved. Therefore a partial--wave decomposition 
of the Schr\"odinger equation results in a system of coupled equations 
for states with different $\ell$. In general this system consists of
an infinite number of equations, and one has to truncate it in 
order to make it tractable. There are, however, certain problems where 
only few partial waves are coupled to each other, namely, those in which 
the non-central part of the potential stems from the nonzero  spins of 
the particles involved. In such systems the total angular momentum,
$$
	      \vec J=\vec\ell+\vec s \, ,
$$
is conserved and the total spin $\vec s$ defines the maximal number 
of coupled partial waves by the triangle condition
\begin{equation}
\label{triangle}
	       |J-s|\le \ell\le |J+s| \, .
\end{equation}
In what follows we will consider a problem of this kind though all 
formulae remain applicable to the more general case of 
coupled partial waves.
\subsection{Partial waves for discrete spectrum}
Wave functions $\Psi_{kJM}({\vec r})$ describing bound and Siegert
(resonance) states are specified by a definite value of the  momentum
$k$, the total angular momentum $J$ and  its third component $M$, and 
the parity $\pi$ which is omitted in our notation. Such wave functions 
can be expanded in terms of the spin--angular functions
$$
	{\cal Y}_{[\ell]}^{JM}({\bf \hat  r})\equiv \sum_{m\mu}\,C_{\ell m
	     s\mu}^{JM} Y_{\ell m}({\bf \hat  r})\chi_{s\mu}
$$
as 
\begin{equation}
\label{partial}
     \Psi_{kJM}(\vec r)=\frac{1}{r}
	\sum_{[\ell]}\,{\cal Y}_{[\ell]}^{JM}(\hat{{\bf r}})
		u_{[\ell]}^J(k,r)\,,
\end{equation}
where $u_{[\ell]}^J(k,r)$ are unknown radial functions, and 
 $[\ell]$ stands for the pair of subscripts $ls$. This notation 
has the advantage that all formulae remain the same for non--central
interactions of spinless particles in which case the symbol 
$[\ell]$  stands for another pair of quantum  numbers, namely,  
$\ell m$ instead of $\ell s$. Furthermore all  formulae can be generalised to  coupled hyperradial equations simply by replacing
$[\ell]$ by $[L]$ where $L$ is the grand orbital quantum number 
(for hyperradial equations see Ref. \cite{fabr}). In what follows we 
drop, for simplicity, the  quantum numbers $JM$ where this does 
not cause a misunderstanding.\\

Substituting  the expansion (\ref{partial}) into the Schr\"odinger
equation, one arrives at the following system of coupled  equations
\begin{equation}
\label{radial}
    \left[\partial^2_r+k^2-\ell(\ell+1)/r^2\right]u_{[\ell]}^J
    (k,r)=\sum_{[\ell']}W_{[\ell][\ell']}^J(r)u_{[\ell']}^J(k,r)\,,
\end{equation}
where the index $[\ell]$ runs over all combinations of $\ell$ and $s$
allowed by the triangle condition (\ref{triangle}) and the parity
conservation law. The elements of the matrix $W$ are those of the operator
$V({\vec r})$ (we use $\hbar$=1),
\begin{equation}
\label{Wmatr}
 	 W_{[\ell][\ell']}^J(r)\equiv 2m\left\langle{\cal Y}_{[\ell]}^{JM}
 	 \right|V({\vec r})\left|{\cal Y}_{[\ell']}^{JM}\right\rangle\,,
\end{equation}
sandwiched between the spin--angular functions (m is the reduced mass).
We assume that these matrix elements are less singular at the origin 
than the centrifugal term,
\begin{equation}
\label{V0}
	\lim_{r\to 0}r^2W_{[\ell][\ell']}^J(r)=0\,,
\end{equation}
and vanish at large distances faster than the Coulomb potential,
\begin{equation}
\label{Vinf}
	\lim_{r\to \infty}rW_{[\ell][\ell']}^J(r)=0\,.
\end{equation}
%
\subsection{Partial waves for continuum}
The scattering state wave function $\Psi_{{\vec k} s\mu}({\vec r})$ is 
defined by the real vector ${\vec k}$ (the momentum of the incoming 
wave), total spin $s$, and its third component $\mu$. The partial wave 
decomposition for this function  is more complicated than for the 
bound and Siegert states because the scattering state depends on 
the direction  of the incident momentum  ${\vec k}$.  Since
$\Psi_{{\vec k} s\mu}({\vec r})$ depends on the two vectors ${\vec k}$ and 
${\vec r}$, we have to perform partial wave analysis in both the momentum 
and coordinate space\cite{newton,gw},
\begin{equation}
\label{scattdecom}
	\Psi_{{\vec k} s\mu}({\vec r})=\sqrt{\frac{2}{\pi}}\frac{1}{kr}
	\sum_{JM[\ell']\ell}\,
	{\cal Y}_{[\ell']}^{JM}(\hat{{\bf r}})u_{[\ell'][\ell]}^J(k,r)
	Y_{[\ell]\mu}^{JM\,*}(\hat{{\bf k}})\, ,
\end{equation}
where
$$
	Y_{[\ell]\mu}^{JM}(\hat{{\bf k}})\equiv
	i^{-\ell}\sum_m C^{JM}_{\ell m s \mu}Y_{\ell m}(\hat{{\bf k}})\, ,
$$
and the radial wave function $u_{[\ell'][\ell]}^J(k,r)$ obeys the same
equation as $u_{[\ell]}^J(k,r)$, 
\begin{equation}
\label{radiall}
    	\left[\partial^2_r+k^2-\ell(\ell+1)/r^2\right]u_{[\ell][\ell']}^J(k,r)
    	=\sum_{[\ell'']}W_{[\ell][\ell'']}^J(r)u_{[\ell''][\ell']}^J(k,r)\, .
\end{equation}
Physical solutions of Eqs. (\ref{radial}) and  (\ref{radiall}) are
defined by the requirement that they must be regular at the
origin,
\begin{equation}
\label{cond0}
	\begin{array}{rc}
	   u_{[\ell]}(k,r)\mathop{\longrightarrow}\limits_{r\to0} 0\, ,& \\
 	   & \\
	   u_{[\ell'][\ell]}(k,r)\mathop{\longrightarrow}\limits_{r\to0}0\,,&
	\end{array}
\end{equation}
and by certain, physically motivated, boundary conditions at infinity,
\begin{equation}
\label{condinf}
	\begin{array}{rc}
	   u_{[\ell]}(k,r)\mathop{\longrightarrow}\limits_{r\to\infty} 
		U_{[\ell]}(k,r)\,,& \\
 	   & \\
	   u_{[\ell'][\ell]}(k,r)\mathop{\longrightarrow}\limits_{r\to\infty}
	   U_{[\ell'][\ell]}(k,r)\,,&
   	\end{array}
\end{equation}
which are different for the various problems (bound, scattering, 
and resonant states) and change drastically when we go over from one case 
to another.
It would be, of course, more convenient  to deal with a universal boundary
conditions imposed at a single point. This can be achieved  if we consider
the general regular solution of Eq. (\ref{radial}) defined only by 
the condition (\ref{cond0}) and not subjected to any restrictions at 
large $r$. This is described next.
\subsection{Regular basis}
From the  vast variety of solutions obeying (\ref{cond0}) and having
different behaviour at large distances we choose only those which are 
linearly independent. They form the fundamental system of regular 
solutions which we call {\it regular basis}. Any regular solution 
with a specific behaviour at  large distances, is simply a linear 
combination of the basic solutions. Thus, instead of having different 
mathematical procedures for the various types of physical problems 
we can have only one for the regular basis. In the next section we 
show how the bound,  scattering, and resonant state wave functions 
can be constructed in terms  of such a basis.\\

Let us consider Eq. (\ref{radial}) as a matrix equation. Then, each of its
solutions is a column--matrix. From the general theory of differential
equations is known that there are as many independent regular
column--solutions of Eq. (\ref{radial}) as the column dimension, i.e., the 
number of equations in the system. These columns can be combined
in a square matrix $\|\Phi_{[\ell][\ell']}\|$  satisfying 
\begin{equation}
\label{regeq}
        \left[\partial^2_r+k^2-\ell(\ell+1)/r^2\right]\Phi_{[\ell][\ell']}
    	(k,r)=\sum_{[\ell'']}W_{[\ell][\ell'']}(r)
    	\Phi_{[\ell''][\ell']}(k,r)\,,
\end{equation}
with
\begin{equation}
\label{zphi}
	\Phi_{[\ell][\ell']}(k,r)\mathop{\longrightarrow}\limits_{r\to0}0\,,
	\qquad \forall\ \ [\ell],[\ell']\,.
\end{equation}
Since Eq. (\ref{regeq}) is of second order and singular at $r=0$, the 
condition (\ref{zphi}) cannot be reduced to the simple requirement
$$
	\Phi_{[\ell][\ell']}(k,0)=0\,,
	\qquad \forall\ \ [\ell],[\ell']\,,
$$
as the behaviour of each element of the matrix $\|\Phi(k,r)\|$ and its 
first derivative $\partial_r\|\Phi(k,r)\|$ 
in the immediate vicinity of the point $r=0$  are also needed.

Like in any other space, the basis can be chosen in an infinite number
of ways by specifying the behaviour (\ref{zphi}).
The possible choice of the condition (\ref{zphi}), however, is not
entirely arbitrary.  It was shown in Ref. \cite{palumbo} (see also an 
alternative proof in the Appendix) that for potentials fulfilling  the 
restriction (\ref{V0}), the regular columns are linearly 
independent only if they vanish near the point $r=0$ in such a way that
\begin{equation}
\label{palum}
	\lim_{r\to 0}\frac{\Phi_{[\ell][\ell']}(k,r)}{r^{\ell'+1}}=
	\delta_{[\ell][\ell']}\, .
\end{equation}
The primed angular momentum in the denominator means that in each row of the 
matrix $\|\Phi_{[\ell][\ell']}\|$ the elements, situated  further to 
the right, vanish faster when $r\rightarrow 0$. Even without a formal proof 
it is clear that such condition guarantees the linear independence of 
the columns since in each row the elements have different behaviour 
(different power of $r$) near $r=0$. If instead of the rows, we look 
at the columns, we get  from Eq. (\ref{palum}) that in each of them 
all off-diagonal elements are infinitesimal as compared to the 
diagonal one.\\

Yet in spite of the restriction (\ref{palum}), we still have some freedom in
specifying the derivatives $\partial_r\Phi_{[\ell][\ell']}$. Indeed,
we can  choose  at least  an  arbitrary  $r$--independent coefficient in each
element of the matrix $\|\Phi_{[\ell][\ell']}\|$. We mention here that it was 
this freedom which was exploited by Newton in his procedure of regularisation 
of the integral equation for $\Phi_{[\ell][\ell']}$ \cite{newton}.\\

To be consistent with the definition of the regular solution for the
un-coupled partial waves \cite{taylor}, we choose the normalisation
coefficients in such a way that
\begin{equation}
\label{fi0}
     	\lim_{r\to 0}\displaystyle{\frac{\Phi_{[\ell][\ell']}(k,r)}
     	{j_{\ell'}(kr)}}=\delta_{[\ell][\ell']}\,,
\end{equation}
where $j_{\ell}(z)$ is the Riccati--Bessel function \cite{abram}.
This condition  defines the leading terms of the near--origin
behaviour of the diagonal elements of the matrix $\|\Phi_{[\ell][\ell']}\|$
and their first derivatives. The off-diagonal elements, however,
remain obscure since the condition (\ref{fi0}) only implies  that 
$\Phi_{[\ell][\ell']}\sim o(j_{\ell'}),  \ell\ne\ell'$.
As pointed out by Newton \cite{newton}, it is impossible to define them
unambiguously by boundary conditions which are independent of the behaviour
of the potential near the origin. In the Appendix we show that this is indeed
the case and we give a simple recipe for obtaining series expansions of all
matrix elements of $\|\Phi(k,r)\|$ at $r\sim0$. Of course, the
terms of such series depend also on the potential.\\

In principle, from the knowledge of such expansions, we could calculate
the matrix $\|\Phi(k,r)\|$ by solving the Schr\"odinger equation
(\ref{regeq}) directly. It is, however, much more convenient to
transform Eq. (\ref{regeq}) into another equivalent form  suitable 
for a construction  of different physical solutions. For this  we 
introduce two new unknown matrices $\|F_{[\ell][\ell']}^{(\pm)}(k,r)\|$ 
and assume the following ansatz for the regular solution
\begin{equation}
\label{ansatz}
	\Phi_{[\ell][\ell']}(k,r)=\frac12\left[
	h_{\ell}^{(+)}(kr)F_{[\ell][\ell']}^{(+)}(k,r)+
	h_{\ell}^{(-)}(kr)F_{[\ell][\ell']}^{(-)}(k,r)\right]
\end{equation}
where the Riccati--Hankel functions $h_{\ell}^{(\pm)}$ are linear
combinations of the Riccati--Bessel and Riccati--Neumann 
functions $h_{\ell}^{(\pm)}(z)\equiv j_{\ell}(z)\pm in_{\ell}(z)$ 
\cite{abram}. The reason for choosing such an ansatz will become 
clear in the next section. Here it is sufficient to say that the 
explicit implantation of the functions $h_{\ell}^{(\pm)}(kr)$ into 
the construction of the basis guarantees the correct asymptotic 
behaviour of the basic solutions at large $r$.\\

Since instead of one unknown matrix $\|\Phi\|$ we introduced two
matrices $\|F^{(\pm)}\|$, they must be subjected to an additional 
constrain. The most convenient is the Lagrange condition
\begin{equation}
\label{lagrange}
	h_\ell^{(+)}(kr)\partial_rF_{[\ell][\ell']}^{(+)}(k,r)+
	h_\ell^{(-)}(kr)\partial_rF_{[\ell][\ell']}^{(-)}(k,r)=0\,,
\end{equation}
which is a standard choice in the variable--constant method for solving  
differential equations \cite{mathews}. Substituting (\ref{ansatz}) into 
Eq. (\ref{regeq}) and using the condition (\ref{lagrange}), we obtain
the following first order coupled differential matrix equations
\begin{equation}
\label{fpmeq}
\partial_rF_{[\ell][\ell']}^{(\pm)}(k,r)=\pm\displaystyle{
	  \phantom{+}\frac{h_\ell^{(\mp)}(kr)}{2ik}
	  \sum_{[\ell'']}}W_{[\ell][\ell'']}(r)\left\{
	  h_{\ell''}^{(+)}(kr)F_{[\ell''][\ell']}^{(+)}(k,r)+
	  h_{\ell''}^{(-)}(kr)F_{[\ell''][\ell']}^{(-)}(k,r)\right\}\,.
\end{equation}
As we show in the Appendix, the boundary condition  (\ref{fi0}) can
be rewritten as
\begin{equation}
\label{fipm0}
     	\lim_{r\to 0}\displaystyle{\left[\frac{
	j_{\ell}(kr)F^{(\pm)}_{[\ell][\ell']}(k,r)}
     	{j_{\ell'}(kr)}\right]}=\delta_{[\ell][\ell']}\,,
\end{equation}
and a series expansion for $ \| F^{(\pm)}_{[\ell][\ell']}(k,r)\|$ near the 
point $r=0$  can be found iteratively as follows
\begin{equation}
\label{recurr0}
	F_{[\ell][\ell']}^{(\pm)(0)}(k,r) = \delta_{[\ell][\ell']}\,,
\end{equation}
\begin{equation}
\label{recurr1}
	     \begin{array}{rcl}
	F_{[\ell][\ell']}^{(\pm)(n+1)}(k,r)&=&\delta_{[\ell][\ell']}\pm
	\displaystyle{\frac{1}{2ik}\int
	h_\ell^{(\mp)}(kr)\sum_{[\ell'']}W_{[\ell][\ell'']}(r)}\\
 & & \\
 	&\times&\left[h_{\ell''}^{(+)}(kr)F_{[\ell][\ell']}^{(+)(n)}(k,r)+
	h_{\ell''}^{(-)}(kr)F_{[\ell][\ell']}^{(-)(n)}(k,r)\right]\,dr\,,
	     \end{array}
\end{equation}
where the arbitrary constant for the indefinite integral is assumed to be zero.
We shall give an explicit example for this expansion in  Section IV.\\

In contrast to Eq. (\ref{regeq}) the equations for the  new unknown matrices
$\|F^{(\pm)}(k,r)\|$ are of first order. However the point  $r=0$ is
generally  a singular point because the Riccati--Hankel functions have the 
short range behaviour $\sim r^{-\ell}$ and according to (\ref{V0}) the 
potential may also behave near this point as $\sim r^{-(2-\varepsilon)}$, 
$\varepsilon>0$. Therefore Eqs. (\ref{fpmeq}) cannot be solved with the 
boundary conditions for $\|F^{(\pm)}(k,r)\|$ at $r=0$. 
Instead,  we may use the analytical solutions of them in a small interval 
$(0,\delta]$ (in the form of the above series expansions) and impose 
the boundary conditions at $r=\delta$.\\

Certain elements of the matrices $\|F^{(\pm)}(k,r)\|$ could diverge when 
$r\rightarrow 0$ (see the Appendix). This, however, does not cause any 
problem in the iterative procedure (\ref{recurr1}) since the integration
is represented by an indefinite integral. It is emphasised that although 
certain elements of $\|F^{(\pm)}\|$ diverge, the matrix $\|\Phi\|$ 
always remains regular because at small $r$ the matrices $\|F^{(\pm)}\|$  
converge to each other, that is they are transformed into some matrix 
$\|A\|$ which is the same for both of them,
\mbox{$\|F^{(\pm)}\|\mathop{\rightarrow}\limits_{r\to 0}\|A\|$}
(see the Appendix), and  the singularities, if any, are compensated by 
the behaviour of the  Riccati--Bessel function $j_\ell(kr)$,
\begin{equation}
\label{compens}
	\Phi_{[\ell][\ell']}(k,r)\mathop{\longrightarrow}\limits_{r\to 0}
	\frac12\left[h_{\ell}^{(+)}(kr)+h_{\ell}^{(-)}(kr)\right]
	A_{[\ell][\ell']}(k,r)=j_\ell(kr)A_{[\ell][\ell']}(k,r)\, .
\end{equation}
The system of equations (\ref{fpmeq}) together with the boundary values
$F_{[\ell][\ell']}^{(\pm)}(k,\delta)\approx
F_{[\ell][\ell']}^{(\pm)(N)}(k,\delta)$ represent a well--defined
differential problem, a solution of which gives the regular basis in the
form (\ref{ansatz}).
\subsection{Jost matrices}
It can be proved (see the Appendix) that for ${\rm Im\,}k=0$ the right 
hand side of Eq.~(\ref{fpmeq}) vanishes when $r\rightarrow\infty$.
Since the corresponding derivatives also vanish, the functions
$F_{[\ell][\ell']}^{(\pm)}(k,r)$ become $r$--independent and
thus, for  momenta corresponding to the scattering
states, we have
\begin{equation}
\label{fiass}
	\Phi_{[\ell][\ell']}(k,r)\mathop{\longrightarrow}\limits_{r\to \infty}
	\frac12\left[h_{\ell}^{(+)}(kr){\cal F}_{[\ell][\ell']}^{(+)}(k)+
	h_{\ell}^{(-)}(kr){\cal F}_{[\ell][\ell']}^{(-)}(k)\right]
\end{equation}
where
\begin{equation}
\label{jplus}
   {\cal F}_{[\ell][\ell']}^{(\pm)}(k)=\lim_{r\to\infty}
	F_{[\ell][\ell']}^{(\pm)}(k,r)\,.
\end{equation}
We may  call these $r$--independent matrices $\|{\cal F}^{(\pm)}\|$ 
as {\it Jost matrices} and the products
$h_{\ell}^{(\pm)}(kr)F_{[\ell][\ell']}^{(\pm)}(k,r)$, which behave
asymptotically like $\sim {\rm e}^{\pm ikr}$,  as {\it Jost solutions}.\\

For complex values of $k$ the above limits generally exist in
different domains of the complex $k$--plane, namely,
${\cal F}_{[\ell][\ell']}^{(+)}(k)$ in the lower half (${\rm Im\,}k\le0$) while
${\cal F}_{[\ell][\ell']}^{(-)}(k)$ in the upper half (${\rm Im\,}k\ge0$).
This is because, according to Eq. (\ref{fpmeq}), the derivatives
$\partial_rF^{(\pm)}(k,r)$ are proportional to $h_\ell^{(\mp)}(kr)$ 
with
\begin{equation}
\label{richaninf}
	h_\ell^{(\pm)}(kr)\mathop{\longrightarrow}\limits_{r\to\infty}
	\mp i\exp\left[\pm i(kr-{\ell\pi/2})\right]
\end{equation}
vanishing  in  different domains of the $k$--plane, namely,
\begin{eqnarray}
\label{rhpls}
	h_\ell^{(+)}(kr)\mathop{\longrightarrow}\limits_{r\to\infty}0\,,
	\qquad {\rm Im\,}(kr)>0\,,\\
\label{rhmns}
	h_\ell^{(-)}(kr)\mathop{\longrightarrow}\limits_{r\to\infty}0\,,
	\qquad {\rm Im\,}(kr)<0\,.
\end{eqnarray}
Thus, in general, the only area where the limits (\ref{jplus}) 
simultaneously exist is the real axis \footnote{
It can also be proved that both these limits exist at all spectral points
corresponding to bound and resonance states (see the Appendix)}.
However, for a particular class of short--range
potentials (decaying exponentially or faster) the upper bound for
the existence of ${\cal F}^{(+)}$ is shifted upwards and the lower bound for
${\cal F}^{(-)}$ downwards, which widens their common area to a band.\\

The difficulty concerning the existence of the limits (\ref{jplus}), can be 
circumvented in the same way as  for central potentials
\cite{nuovocim,nth9607028}. Indeed, the conditions (\ref{rhpls}) 
and (\ref{rhmns}) involve the imaginary part of the product $kr$ and not 
of the momentum alone. Therefore, if, for example, ${\rm Im\,}(kr)$ 
is negative we can make it positive by using the complex rotation method 
which we describe next.
\subsection{Complex rotation}
In this method the radius $r$ is replaced by a 
complex one, viz., 
\begin{equation}
\label{rot}
	r=x\exp(i\theta)\, ,\qquad x\ge 0\, ,
		\qquad 0\le|\theta|<\frac{\pi}{2}\, .
\end{equation}
The idea of complex rotation of the coordinate is not new. Many years ago, 
during World War II, Hartree and his co-workers at Manchester university 
used such rotation to solve certain differential equations describing 
radio wave propagation in the atmosphere (for more  details
see Ref.\cite{connor}). Nowadays the complex rotation is widely used for
locating quantum resonances by variational methods \cite{ccr}. In contrast,
our equations together with the complex rotation can be used to locate
resonances in an exact way.\\

Applying the complex transformation (\ref{rot}) to Eq. (\ref{fpmeq}), we
obtain
\begin{equation}
\label{fpmrot}
	   \begin{array}{l}
		\partial_xF_{[\ell][\ell']}^{(\pm)}(k,x{\rm e}^{i\theta})
		=\pm\displaystyle{\phantom{+}\frac{{\rm e}^{i\theta}
		h_\ell^{(\mp)}(kx{\rm e}^{i\theta})}{2ik}
	  	\sum_{[\ell'']}}W_{[\ell][\ell'']}(x{\rm e}^{i\theta})\\
		\phantom{-}\\\phantom{-------}\times
	\left\{
		h_{\ell''}^{(+)}(kx{\rm e}^{i\theta})F_{[\ell'']
		[\ell']}^{(+)}(k,x{\rm e}^{i\theta})+
		h_{\ell''}^{(-)}(kx{\rm e}^{i\theta})
		F_{[\ell''][\ell']}^{(-)}(k,x{\rm e}^{i\theta})
	\right\}\,.\\
	   \end{array}
\end{equation}
The purpose of the rotation (\ref{rot}) is to make the imaginary part of the
product $kr$  positive or negative, in calculating ${\cal F}^{(-)}$ or
${\cal F}^{(+)}$ respectively, at points on the $k$--plane
we are interested in. Thus we have ${\rm Im\,}kr>0$  for all points above
the dividing line shown on Fig. \ref{rotfig}. This line defined by the 
negative angle $\theta$ in the $k$--plane results from  the rotation 
(\ref{rot}) in the $r$--plane with  positive $\theta$.

If the potential matrix $\|W(r)\|$ is an analytic function of the 
complex variable $r$ and obeys the conditions (\ref{V0}, \ref{Vinf}) 
along the ray (\ref{rot}), then the limit
\begin{equation}
\label{fmlim}
	\lim_{x\to\infty}F_{[\ell][\ell']}^{(-)}(k,x{\rm e}^{i\theta})=
	{\cal F}_{[\ell][\ell']}^{(-)}(k)
\end{equation}
exists and is finite for all $k$ on and above the dividing line
$[-\infty {\rm e}^{-i\theta},+\infty {\rm e}^{-i\theta}]$ (for the relevant proof of this
statement see the Appendix). At the same time the limit
\begin{equation}
\label{fplim}
	\lim_{x\to\infty}F_{[\ell][\ell']}^{(+)}(k,x{\rm e}^{i\theta})=
	{\cal F}_{[\ell][\ell']}^{(+)}(k)
\end{equation}
exists and is finite for all $k$ on and below the dividing line. Moreover,
when the limits (\ref{fmlim}) and  (\ref{fplim}) exist the values of
${\cal F}_{[\ell][\ell']}^{(\pm)}(k)$ are independent of the rotation angle
$\theta$ as the Jost function is $r$--independent and hence 
$\theta$--independent. Thus, the limits (\ref{fmlim}) and (\ref{fplim}) 
give a  unique analytic continuation of the Jost matrices to the lower 
and upper halves of the complex $k$--plane respectively. To calculate the
${\cal F}_{[\ell][\ell']}^{(-)}(k)$ for ${\rm Im\,}k<0$ we need to solve
Eqs. (\ref{fpmrot}) at a sufficiently large positive $\theta$, and the
${\cal F}_{[\ell][\ell']}^{(+)}(k)$ for ${\rm Im\,}k>0$  at a sufficiently 
large negative $\theta$.\\

We note that though the ansatz
(\ref{ansatz}) is suitable for large distances (see forth the
next section), it is not good for numerical calculations in the
vicinity of $r=0$. Indeed, near this point the singularities of 
$h_\ell^{(+)}(kr)$ and $h_\ell^{(-)}(kr)$ are cancelled. Although 
this does not cause formally any problem, in numerical calculations
the cancellation of singularities is always a source of possible 
numerical errors. These errors increase with
increasing $\ell$ since in this case $h_\ell^{(\pm)}(kr)$ becomes more
singular. Therefore, for larger $\ell$ the point $r=\delta$ must
be shifted further from the origin. This in turn, requires more iterations 
of Eq. (\ref{recurr1}) to obtain the boundary values $F_{\ell\ell'}^{(\pm)}
(k,\delta)$ to a required accuracy.\\

Eq. (\ref{compens}) hints to another way to handle numerically the 
boundary condition problem. Since
$$
	\frac12(h_\ell^++h_\ell^-)=j_\ell\qquad{\rm and}\qquad
	\frac{1}{2i}(h_\ell^+-h_\ell^-)=n_\ell\, ,
$$
we may introduce a new pair of matrices
\begin{eqnarray}
\nonumber
	A_{[\ell][\ell']}(k,r)\equiv\frac{1}{2}\left[
	F_{[\ell][\ell']}^{(+)}(k,r)
	+F_{[\ell][\ell']}^{(-)}(k,r)\right]\,,\\
\label{AB} \\
\nonumber
	B_{[\ell][\ell']}(k,r)\equiv\frac{1}{2i}\left[
	F_{[\ell][\ell']}^{(+)}(k,r)
	-F_{[\ell][\ell']}^{(-)}(k,r)\right]\,,
\end{eqnarray}
which transform the ansatz (\ref{ansatz}) into the form
\begin{equation}
\label{newans}
	\Phi_{[\ell][\ell']}(k,r)=j_\ell(kr)A_{[\ell][\ell']}(k,r)
	-n_\ell(kr)B_{[\ell][\ell']}(k,r)\,,
\end{equation}
and the corresponding linear combination of equations (\ref{fpmeq}) gives the
alternative form 
\begin{equation}
\label{ABeq}
\left\{
\begin{array}{l}
\partial_rA_{[\ell][\ell']}(k,r)=\displaystyle{
	  -\frac{n_\ell(kr)}{k}
	  \sum_{[\ell'']}}W_{[\ell][\ell'']}(r)\left[
	  j_{\ell''}(kr)A_{[\ell''][\ell']}(k,r)-
	  n_{\ell''}(kr)B_{[\ell''][\ell']}(k,r)\right]\,,\\
\phantom{-}\\
\partial_rB_{[\ell][\ell']}(k,r)=\displaystyle{
	  -\frac{j_\ell(kr)}{k}
	  \sum_{[\ell'']}}W_{[\ell][\ell'']}(r)\left[
	  j_{\ell''}(kr)A_{[\ell''][\ell']}(k,r)-
	  n_{\ell''}(kr)B_{[\ell''][\ell']}(k,r)\right]\,.
\end{array}
\right.
\end{equation}
Likewise, the iterative procedure (\ref{recurr0}, \ref{recurr1}) transforms
into 
\begin{equation}
\label{A0B0}
	A_{[\ell][\ell']}^{(0)}(k,r)=\delta_{[\ell][\ell']}\,,\qquad
	B_{[\ell][\ell']}^{(0)}(k,r)=0\,,
\end{equation}
\begin{equation}
\label{ABiter}
\left\{
\begin{array}{rcl}
A_{[\ell][\ell']}^{(n+1)}(k,r)&=&\delta_{[\ell][\ell']}-
	  \displaystyle{\frac{1}{k}\int n_\ell(kr)
	  \sum_{[\ell'']}}W_{[\ell][\ell'']}(r)\\
 & & \phantom{-}\\
	  &\times&\left[j_{\ell''}(kr)A_{[\ell''][\ell']}^{(n)}(k,r)-
	  n_{\ell''}(kr)B_{[\ell''][\ell']}^{(n)}(k,r)\right]\,dr\, ,\\
 & & \phantom{-}\\
B_{[\ell][\ell']}^{(n+1)}(k,r)&=&
	  -\displaystyle{\frac{1}{k}\int j_\ell(kr)
	  \sum_{[\ell'']}}W_{[\ell][\ell'']}(r)\left[
	  j_{\ell''}(kr)A_{[\ell''][\ell']}^{(n)}(k,r)-
	  n_{\ell''}(kr)B_{[\ell''][\ell']}^{(n)}(k,r)\right]\,dr\, .
\end{array}
\right.
\end{equation}
The representation of $\|\Phi\|$ in terms of $\|A\|$, $\|B\|$
and $\|F^{(\pm)}\|$ is equivalent. From a  practical point of view,
however, it is more convenient to start the integration of Eqs. (\ref{ABeq})
from the boundary values $A_{[\ell][\ell']}(k,\delta)$,
$B_{[\ell][\ell']}(k,\delta)$ and at some intermediate point $r_{int}$ (far
enough from the origin) to go over to Eqs. (\ref{fpmeq}) with starting
values $F_{[\ell][\ell']}^{(\pm)}(k,r_{int})$ obtained from
$A_{[\ell][\ell']}(k,r_{int})$ and $B_{[\ell][\ell']}(k,r_{int})$ via the
linear combinations (\ref{AB}).\\

One may argue that we can abandon equations (\ref{fpmeq}) altogether and
integrate instead Eq. (\ref{ABeq}) on the whole interval
$[\delta,r_{max}]$. However, $\|F^{(+)}(k,r)\|$ and $\|F^{(-)}(k,r)\|$ 
have finite limits $(r\to\infty)$ in different domains of the complex 
$k$--plane (below and above the dividing line respectively). The only 
points where they have limits simultaneously are the spectral points 
and the dividing line itself. And since $\|A\|$ and $\|B\|$ involve both 
$\|F^{(\pm)}\|$, they have limits only at these points. Therefore, to 
obtain the Jost matrix we should start at small $r$ with Eq. (\ref{ABeq}) 
and finish at  large $r_{max}$ with Eq. (\ref{fpmeq}).
%
\section{Physical solutions}
In what follows we shall describe how  we can obtain a physical solution 
from the regular basis. In general, each column representing a physical 
solution is a linear combination of the basic columns,
\begin{equation}
\label{lincomb}
		\begin{array}{rcl}
	u_{[\ell]}(k,r)&=&
	\displaystyle{\sum_{[\ell']}}\Phi_{[\ell][\ell']}(k,r)c_{[\ell']}\,,\\\\
	u_{[\ell'][\ell]}(k,r)&=&
	\displaystyle{\sum_{[\ell'']}}\Phi_{[\ell'][\ell'']}(k,r)
	c_{[\ell''][\ell]}\,,
		\end{array}
\end{equation}
with the coefficients $\|c\|$  defined by the physical
boundary condition (\ref{condinf}) at large distances.
\subsection{Bound states}
The bound state wave function vanishes at large distances as
$$
	\sum_{[\ell']}\Phi_{[\ell][\ell']}(k,r)c_{[\ell']}
	\mathop{\longrightarrow}\limits_{r\to \infty}N_{[\ell]}{\rm e}^{-|k|r}
	\mathop{\longrightarrow}\limits_{r\to \infty}0\, ,
$$
where $N_{[\ell]}$ are the asymptotic normalisation constants.
In this equation the function $\Phi_{[\ell][\ell']}(k,r)$ can 
be replaced by its asymptotic form (\ref{fiass}), i.e.,
\begin{equation}
\label{bcond}
	\frac12\sum_{[\ell']}\left[h_{\ell}^{(+)}(kr){\cal F}_{[\ell][\ell']}^{(+)}(k)+
	h_{\ell}^{(-)}(kr){\cal F}_{[\ell][\ell']}^{(-)}(k)\right]c_{[\ell']}
	\mathop{\longrightarrow}\limits_{r\to \infty}0\, .
\end{equation}
For bound states ${\rm Im\,}k>0$ and the Riccati--Hankel
function $h_\ell^{(+)}(kr)$ decays exponentially while $h_\ell^{(-)}(kr)$
grows exponentially. Therefore the condition (\ref{bcond}) can be fulfilled
only if we find  coefficients $c_{[\ell]}$ such that the diverging functions
$h_\ell^{(-)}(kr)$ of different columns cancel out, that is, if
\begin{equation}
\label{homo}
	\sum_{[\ell']}{\cal F}_{[\ell][\ell']}^{(-)}(k)c_{[\ell']}=0\,.
\end{equation}
This system of homogeneous linear equations has a nontrivial solution if and
only if
\begin{equation}
\label{det0}
	\det\|{\cal F}^{(-)}(k)\|=0\,.
\end{equation}
Therefore, we can locate all possible bound states by looking for zeros $k_0$
of the Jost--matrix determinant on the positive imaginary axis (see Fig. 
\ref{rotfig}). For each zero $k_0$ thus found, the coefficients $c_{[\ell]}$
are then uniquely determined by the system (\ref{homo}) apart from a
general normalisation factor which is finally fixed when the physical
wave function,
\begin{equation}
\label{psibound}
	\Psi_{k_0JM}({\vec  r})=\frac{1}{2r}\sum_{[\ell][\ell']}
	{\cal Y}_{[\ell]}^{JM}(\hat{{\bf r}})
	\left[h_{\ell}^{(+)}(k_0r)F_{[\ell][\ell']}^{(+)}(k_0,r)+
	h_{\ell}^{(-)}(k_0r)F_{[\ell][\ell']}^{(-)}(k_0,r)\right]c_{[\ell']}\,,
\end{equation}
is normalised. The contribution of each element of the column
\begin{equation}
	u_{[\ell]}(k_0,r)=\frac{1}{2}\sum_{[\ell']}
	\left[h_{\ell}^{(+)}(k_0r)F_{[\ell][\ell']}^{(+)}(k_0,r)+
	h_{\ell}^{(-)}(k_0r)F_{[\ell][\ell']}^{(-)}(k_0,r)\right]c_{[\ell']}
\end{equation}
into the normalisation integral represents what is usually called the
percentage of the corresponding partial wave.
\subsection{Scattering states}
The scattering states normalised to the $\delta$--function,
$$
	\langle\Psi_{{\vec k} s\mu}|\Psi_{{\vec k}' s'\mu'}\rangle=
	\delta({\vec k} - {\vec k}')\delta_{ss'}\delta_{\mu\mu'} \,,
$$
are defined by the following asymptotic condition
\begin{equation}
\label{scatt3}
	\Psi_{{\vec k}s\mu}({\vec r})\,\mathop{\longrightarrow}_{r\to\infty}\,
	\frac{1}{(2\pi)^{3/2}}\left[{\rm e}^{i{\vec  k}\cdot{\vec r}}
	\chi_{s\mu}+\frac{{\rm e}^{ikr}}{r}\sum_{s'\mu'}
	{\rm f}_{s'\mu' s\mu}(k\frac{{\vec r}}{r},{\vec k})\chi_{s'\mu'}
         \right]\, ,
\end{equation}
where ${\rm f}_{s'\mu' s\mu}({\vec k}',{\vec k})$ is the scattering amplitude.
The partial wave decomposition of Eq. (\ref{scatt3}) gives for the boundary 
condition (\ref{condinf}) (see Ref.\cite{newton})
\begin{equation}
\label{scattr}
	u_{[\ell'][\ell]}(k,r)\,\mathop{\longrightarrow}_{r\to\infty}\,
	\frac12\left[h^{(-)}_{\ell'}(kr)\delta_{[\ell'][\ell]}
	+h^{(+)}_{\ell'}(kr)S_{[\ell'][\ell]}^J(k)\right]\, .
\end{equation}
Therefore the choice of  our ansatz (\ref{ansatz}) for the regular basis
is natural and suitable not only for constructing the bound states but
the scattering states as well. Indeed, comparing (\ref{scattr}) with
(\ref{fiass}) we find that the coefficients
$c_{[\ell'][\ell]}$ in (\ref{lincomb}) should be chosen as
$$
	c_{[\ell'][\ell]}=\|{\cal F}_{[\ell'][\ell]}^{(-)}(k)\|^{-1}\,,
$$
which gives us the $S$--matrix in the form
\begin{equation}
\label{smatr}
	\|S(k)\|=\|{\cal F}^{(+)}(k)\|\cdot\|{\cal F}^{(-)}(k)\|^{-1}\, .
\end{equation}
Thus, the normalised scattering wave function  can be constructed
from the regular basis as follows
\begin{eqnarray}
\label{scattwf}
	\Psi_{{\vec k} s\mu}({\vec r})&=&\displaystyle{\frac{1}{\sqrt{2\pi}kr}
	\sum_{JM\ell}\sum_{[\ell']}\sum_{[\ell'']}}\\
\nonumber
	&\phantom{+}&
	{\cal Y}_{[\ell']}^{JM}(\hat{{\bf r}})Y_{[\ell]\mu}^{JM\,*}
	(\hat{{\bf k}}) \left\{
	h_{\ell'}^{(+)}(kr)F_{[\ell'][\ell'']}^{(+)}(k,r)+
	h_{\ell'}^{(-)}(kr)F_{[\ell'][\ell'']}^{(-)}(k,r)\right\}
	\|{\cal F}_{[\ell''][\ell]}^{(-)}(k)\|^{-1}\,.
\end{eqnarray}
The scattering phase shifts together with the mixing parameters can be
found from the $S$--matrix given by Eq. (\ref{smatr}).\\
\subsection{Resonances}
The resonance (or Siegert) states are described by wave functions which
at large distances have only outgoing spherical waves
\begin{equation}
\label{siegert}
	\Psi_{kJM}({\vec r})=\frac{1}{r}
	\sum_{[\ell]}\,{\cal Y}_{[\ell]}^{JM}(\hat{{\bf r}})u_{[\ell]}^J(k,r)
	\,\mathop{\longrightarrow}_{r\to\infty}\,
	\sim\frac{{\rm e}^{ikr}}{r}\,.
\end{equation}
More precisely, the resonant boundary condition at large $r$ is
\begin{equation}
\label{reson}
	u_{[\ell]}^J(k,r)\,\mathop{\longrightarrow}_{r\to\infty}\,
	N_{[\ell]}^J(k)h_{\ell}^{(+)}(kr)\,,
\end{equation}
where $N_{[\ell]}^J(k)$ are the partial wave normalisation constants.
Since there is not a generally accepted convention about the normalisation of
Siegert states, the  $N_{[\ell]}^J(k)$ may involve an arbitrary
coefficient. Solutions of the Schr\"odinger equation, with the long--range
behaviour (\ref{reson}), may exist  only at discrete points of the complex
$k$--plane, situated below the real axis (see Fig. \ref{rotfig}).
The corresponding radial wave functions are regular at
the origin and are therefore linear combinations of the regular basis
\begin{equation}
\label{siecomb}
	u_{[\ell]}(k,r)=\frac12\sum_{[\ell']}
	\left\{h_{\ell}^{(+)}(kr)F_{[\ell][\ell']}^{(+)}(k,r)+
	h_{\ell}^{(-)}(kr)F_{[\ell][\ell']}^{(-)}(k,r)\right\}c_{[\ell']}\, .
\end{equation}
This equation can be used  to locate resonances. Similarly to 
the bound states, we  simply require  that
\begin{equation}
\label{limsie}
	\lim_{r\to\infty}\sum_{[\ell']}F_{[\ell][\ell']}^{(-)}
	(k,x{\rm e}^{i\theta})
	c_{[\ell']}=0\,,\qquad \theta>0\, ,
\end{equation}
which means that $u_{[\ell]}(k,r)$  is expressed only in terms of
$h_\ell^{(+)}(k,r)$  at large distances. Since $\theta$ can be chosen 
such that ${\rm Im\,}kr>0$, the Riccati--Hankel function $h_{\ell}^{(+)}(kr)$ 
decays exponentially when $|r|\to \infty$. This in turn means that the Siegert 
states are quadratically integrable according to (\ref{reson}) and hence 
they acquire the same properties as the bound states. Therefore, with  the 
coordinate rotation the bound and resonance states can be treated in the
same way. In particular, the position of a resonance is defined by 
Eq. (\ref{det0}), and its (rotated) wave function by Eq. (\ref{psibound}) 
where $r$ is now complex.\\

Since this wave function is square integrable we can normalise it to unity
and thus  the normalisation constants $N_{[\ell]}^J$ in (\ref{reson}) 
can be fixed in a natural way. Therefore the proposed method enables us   
not only to locate the position of the resonances as zeros of the Jost 
matrix determinant but also to obtain the correct normalisation constants 
$N_{[\ell]}^J$. In order to obtain physical (unrotated) Siegert wave 
functions, the Schr\"odinger equation must 
be integrated along real $r$ inwards  using the boundary condition 
(\ref{reson}), the  found momentum $k_0$ and normalisation constants. 
This integration will provide  automatically
a wave function which is zero at $r=0$ because $k_0$ is a spectral point. 
\section{Examples}
In order to demonstrate the effectiveness of the method we consider as as 
example the nucleon--nucleon (NN) interaction in the triplet spin--state, 
i.e., when the total spin $s=1$.  From the Pauli principle it follows that 
$s$ is conserved (see for example Ref.\cite{landau}), and thus the sum over
$[\ell]$ is reduced to $\sum_\ell$. The triplet NN potential can couple
only at most two partial waves, with $\ell=J-1$ and $\ell=J+1$, 
as the state with $\ell=J$ has different parity and therefore 
must be excluded. We consider here the even
state of two nucleons with $J=1$, in which they can form the deuteron. 
The partial wave decomposition of this state consists of coupled
$S$ and $D$ waves.\\

The corresponding NN potential includes the following three most important
terms
\begin{equation}
\label{NNpot}
	V({\vec r})=V_c(r)+V_t(r)S_{12}+V_{\ell s}(r)(\vec\ell\cdot\vec s)\,,
\end{equation}
known as the central, tensor, and spin--orbit potentials. The second term
contains the tensor operator
$$
	S_{12}=\frac{3}{r^2}(\vec\sigma_1\cdot\vec r)
	  (\vec\sigma_2\cdot\vec r)-(\vec\sigma_1\cdot\vec\sigma_2)
$$
which is responsible for the coupling of different partial waves. This is clearly
seen from the structure of the matrix $\|W\|$ defined by Eq. (\ref{Wmatr}),
which in the case of $s=1$, $J=1$, and $\pi=+1$ reads (see for example Ref.
\cite{newton})
\begin{equation}
\label{Wstruct}
	\left(
	  \begin{array}{cc}
		W_{00}     &  W_{02} \\
 		           &         \\
		W_{20}     &   W_{22}
	\end{array}
	\right)\equiv 2m\left(
	   \begin{array}{cc}
                    V_c    &\    2\sqrt{2}V_t     \\
 			   &      		 \\
		2\sqrt{2}V_t  &\ V_c-2V_t-3V_{\ell s}
	\end{array}
\right)\,,
\end{equation}
where the subscripts of $\|W_{\ell\ell'}\|$ correspond to the two partial
waves $\ell=0$ and $\ell=2$.\\

In accordance with the restriction (\ref{V0}) we may use at short 
distances  the series expansion 
\begin{equation}
\label{wzero}
	W_{\ell\ell'}(r)\,\mathop{\longrightarrow}\limits_{r\to 0}\,
	a_{\ell\ell'}r^{-1}+b_{\ell\ell'}+c_{\ell\ell'}r+\dots\, ,
\end{equation}
which is general enough to include potentials having a soft core 
($ a_{\ell\ell'}\ne 0$).\\

Using (\ref{wzero}) we can obtain the corresponding expansions for 
$A_{\ell\ell'}(k,r)$  and $B_{\ell\ell'}(k,r)$ via the iterative procedure 
(\ref{A0B0}, \ref{ABiter}) which are needed to  impose the boundary conditions
$A_{\ell\ell'}(k,\delta)$  and $B_{\ell\ell'}(k,\delta)$  at a small
$\delta$. These expansions near the point $r=0$  can be obtained in
an explicit form. For this  we replace the Riccati functions in the
indefinite integrals (\ref{ABiter}) by their series expansions \cite{abram}
\begin{eqnarray}
\nonumber
	j_0(kr)&=&kr-\frac{(kr)^3}{6}+\frac{(kr)^5}{120}-\cdots\,,\\
\nonumber
	j_2(kr)&=&
       \frac{(kr)^3}{15}-\frac{(kr)^5}{210}+\frac{(kr)^7}{7560}-\cdots\,,\\
\nonumber
	n_0(kr)&=&-1+\frac{(kr)^2}{2}-\frac{(kr)^4}{24}+\cdots\,,\\
\nonumber
	n_2(kr)&=&-\frac{3}{(kr)^2}-\frac12-\frac{(kr)^2}{8}-\cdots\,.
\end{eqnarray}
Substituting (\ref{A0B0}) into (\ref{ABiter}) and performing
the integrations over $r$ for each element of the matrices $\|A\|$ and
$\|B\|$ we obtain
$$
	\|A^{(1)}(k,r)\|=
	   \left(
	\begin{array}{cc}
     \displaystyle{1+a_{00}r}    &  \  \displaystyle{\frac{a_{02}k^2}{45}r^3}\\
 	        		 &                                          \\
     \displaystyle{-\frac{3a_{20}}{k^2}r^{-1}}   &\
                                     \displaystyle{1+\frac{a_{22}}{5}r}
	\end{array}
	\right)\,,\qquad
	\|B^{(1)}(k,r)\|=
	\left(
	\begin{array}{cc}
     \displaystyle{-\frac{a_{00}k}{2}r^2}      &\
		\displaystyle{-\frac{a_{02}k^3}{60}r^4}\\
 				& \\
		\displaystyle{-\frac{a_{20}k^3}{60}r^4}   &\
		\displaystyle{-\frac{a_{22}k^5}{1350}r^6}
	 \end{array}
	   \right)\,,
$$
where we retain only one additional term in each matrix element. 
For the second iteration  we get
$$
	\|A^{(2)}(k,r)\|=
	\left(
	\begin{array}{cc}
	\displaystyle{1+a_{00}r+\frac{\zeta_{00}}{2}r^2}
	&\ \displaystyle{\frac{a_{02}k^2}{45}r^3+\frac{k^2\zeta_{02}}{60}r^4}\\
 		&  \\
	\displaystyle{-\frac{3a_{20}}{k^2}r^{-1}+\frac{3\zeta_{20}}{k^2}\ln r}
	&\ \displaystyle{1+\frac{a_{22}}{5}r+\frac{\zeta_{22}}{10}r^2}
\end{array}
\right)\,,
$$
$$
	\|B^{(2)}(k,r)\|=
	\left(
	\begin{array}{cc}
	\displaystyle{-\frac{a_{00}k}{2}r^2-\frac{k\zeta_{00}}{3}r^3}&\
	\displaystyle{-\frac{a_{02}k^3}{60}r^4-\frac{k^3\zeta_{02}}{75}r^5}\\
   		& \\
	\displaystyle{-\frac{a_{20}k^3}{60}r^4-
	\frac{k^3\zeta_{20}}{75}r^5} &\
	\displaystyle{-\frac{a_{22}k^5}{1350}r^6-
	\frac{k^5\zeta_{22}}{1575}r^7}
	\end{array}
	\right)\,,
$$
where the constants $\zeta_{\ell\ell'}$ are defined as
$$
	\left(
	\begin{array}{cc}
	\zeta_{00} & \zeta_{02} \\
 	 & \\
	\zeta_{20} & \zeta_{22}
	\end{array}
		\right) = \left(
	\begin{array}{cc}
	\displaystyle{\frac{a_{00}^2}{2}+b_{00}-\frac{a_{02}a_{20}}{4}}&\
	\displaystyle{\frac{a_{00}a_{02}}{12}+b_{02}+\frac{a_{02}a_{22}}{6}}\\
 		& \\
	\displaystyle{\frac{a_{20}a_{00}}{2}+b_{20}-\frac{a_{22}a_{20}}{4}} &\
	\displaystyle{\frac{a_{20}a_{02}}{12}+b_{22}+\frac{a_{22}^2}{6}}
	\end{array}
	\right)\,.
$$
The iterations can be continued in the same manner with each
iteration adding a new term to each matrix element having
a higher power of $r$ than the previous one. All matrix
elements of $\|A\|$ and $\|B\|$ are regular when $r\to 0$ except for the
left bottom corner element of $\|A\|$. It has two singular terms,
$\sim r^{-1}$ and $\sim\ln r$. The next iteration however gives for it a
vanishing term of the kind $\sim r\ln r$, and all subsequent terms are
vanishing even faster.\\

The above  expansions illustrate the fact that at small $r$ the matrices
$\|F^{(+)}\|$ and $\|F^{(-)}\|$ converge to each other. Indeed, since $\|B\|$ 
is infinitesimal as compared to $\|A\|$ the second term in the linear
combinations $\|F^{(\pm)}\|=\|A\|\pm i\|B\|$ vanishes when $r\to 0$.\\

The singularities of the matrix $\|A_{[\ell][\ell']}(k,r)\|$ reflects the 
main difficulty of the theory of coupled partial waves, which precluded its 
development in the past. As we show in the Appendix, the matrix elements of
$\|F_{[\ell][\ell']}^{(\pm)}(k,r)\|$ with $\ell>\ell'$ are always singular
at $r=0$ if the potential is nonzero at that point.
Despite this the functions $\Phi_{[\ell][\ell']}(k,r)$ remain always regular
because in Eq. (\ref{newans}) the singularity of $A_{[\ell][\ell']}(k,r)$
is compensated by $j_\ell(kr)$. In our case we have
\begin{equation}
\label{phjanb}
	\|\Phi\|=\left(
	\begin{array}{ccc}
		j_0 & \phantom{,} & 0 \\
		0 & \phantom{,} &  j_2
	\end{array}
	\right)\left(
	\begin{array}{ccc}
		A_{00} & \phantom{,} & A_{02} \\
		A_{20} & \phantom{,} & A_{22}
	\end{array}
	\right)-\left(
	\begin{array}{ccc}
		n_0 & \phantom{,} & 0 \\
		0 & \phantom{,} &  n_2
	\end{array}
	\right)\left(
	\begin{array}{ccc}
		B_{00} & \phantom{,} & B_{02} \\
		B_{20} & \phantom{,} & B_{22}
	\end{array}
	\right)
\end{equation}
and therefore the singular term $A_{20}(k,r)$ appears only in the product
$j_2(kr)A_{20}(k,r)$ which vanishes at the origin as $\sim r^2$. Substituting 
the series expansions of all functions involved into Eq. (\ref{phjanb}) 
we obtain
\begin{equation}
\label{phser}
	\begin{array}{l}
	\|\Phi(k,r)\|\mathop{\longrightarrow}\limits_{r\to 0}\\ \\
		\left(
		\begin{array}{cc}
	kr+\displaystyle{\frac{a_{00}k}{2}r^2+\frac{\zeta_{00}k}{6}r^3}+
	{\cal O}(r^4\ln r)  &\ 
	\displaystyle{\frac{a_{02}k^3}{180}r^4+\frac{\zeta_{02}k^3}{300}r^5}+
	{\cal O}(r^6)\\
 		&  \\
	\displaystyle{-\frac{a_{20}k}{4}r^2+\frac{\zeta_{20}k}{5}r^3\ln r -
	\frac{\zeta_{20}k}{25}r^3}+{\cal O}(r^4\ln r) &\
	\displaystyle{\frac{k^3}{15}r^3+\frac{a_{22}k^3}{90}r^4+
	\frac{\zeta_{22}k^3}{210}r^5}+{\cal O}(r^6)
	\end{array}
		\right)
		\end{array}
\end{equation}
which  explicitly  demonstrates that our regular basis obeys the boundary
condition (\ref{fi0}) and, as mentioned in Sec. II, in each row the 
convergence to zero increases from left to right while in each column
the diagonal elements have the lowest vanishing speed. The linear
independence of these columns is also apparent. As far as the off-diagonal 
elements are concern, even the leading terms of them depend on the behaviour 
of the potential, i.e., it is impossible to specify the boundary condition 
for them in a general form independently from the potential.\\

Using a quite different approach, Palumbo in Ref. \cite{palumbo} derived
recurrence formulae for direct constructing the series expansion of the
regular basis for the potentials of the type (\ref{wzero}). Having performed
few iterations of the Palumbo's formulae, we found that they generate the
same terms which are given in Eq. (\ref{phser}). This is yet another
confirmation  that our iterative procedures (\ref{recurr0}, 
\ref{recurr1}) and (\ref{A0B0}, \ref{ABiter}) are correct.\\

In order to test the proposed method numerically, we chose two different
NN--potentials of the type (\ref{wzero}). The first one is the
Reid soft core potential (RSC) \cite{RSC}, which for $J=1$, $s=1$,
$\pi=+1$ has the following form
\begin{equation}
\label{RSCpot}
		\begin{array}{lcl}
	V_c(r)&=&\displaystyle{ h_0\frac{{\rm e}^{-\alpha r}}{\alpha r}+
	h_1\frac{{\rm e}^{-2\alpha r}}{\alpha r}+
	h_2\frac{{\rm e}^{-4\alpha r}}{\alpha r}+
	h_3\frac{{\rm e}^{-6\alpha r}}{\alpha r}\,,}\\
 	& & \\
	V_t(r)&=&\displaystyle{
	h_0\left\{\left[\frac{1}{\alpha r}+\frac{3}{(\alpha r)^2}+
	\frac{3}{(\alpha r)^3}\right]{\rm e}^{-\alpha r}
	-\left[\frac{12}{(\alpha r)^2}+\frac{3}{(\alpha r)^3}\right]
	{\rm e}^{-4\alpha r}\right\}+   }\\
 	& & \\
	& &\displaystyle{+
	h_4\frac{{\rm e}^{-4\alpha r}}{\alpha r}+h_5\frac{{\rm e}^{-6\alpha r}}
	{\alpha r}\,,}\\
 	& & \\
	V_{\ell s}(r)&=&\displaystyle{
	h_6\frac{{\rm e}^{-4\alpha r}}{\alpha r}+
	h_7\frac{{\rm e}^{-6\alpha r}}{\alpha r}\,,}
		\end{array}
\end{equation}
with
$$
	\begin{array}{lll}
	h_0=-10.463\, {\rm MeV}\,,&\quad h_1=105.468\, {\rm MeV}\,,&\quad h_2
	=-3187.8\, {\rm MeV}\,,\\
	h_3=9924.3\, {\rm MeV}\,,&\quad h_4=351.77\, {\rm MeV}\,,&\quad h_5
		=-1673.5\, {\rm MeV}\,,\\
	h_6=708.91\, {\rm MeV}\,,&\quad h_7=-2713.1\,{\rm MeV}\,,&\quad 
	\alpha	=0.7\, {\rm fm}^{-1}\,.
	\end{array}
$$
The second potential used, is the Moscow potential \cite{moscow},
\begin{eqnarray}
\nonumber
	V_c(r)&=&V_1{\rm e}^{-\eta r^2}+V_2\left(1-{\rm e}^{-\gamma r}\right)
	\frac{{\rm e}^{-\beta r}}{\beta r}\,,\\
\label{moscowpot}
	V_t(r)&=&V_2\left[1+\frac{3}{\beta r}+\frac{3}{(\beta r)^2}\right]
	\left(1-{\rm e}^{-\gamma r}\right)^3\frac{{\rm e}^{-\beta r}}{\beta r}\,,\\
\nonumber
	V_{\ell s}(r) &\equiv &0\,,
\end{eqnarray}
with
$$
\begin{array}{lll}
	V_1=-466.74\, {\rm MeV}\,,&\quad  V_2=-10.69\, {\rm MeV}\,,\quad & \\
	\beta=0.6995\, {\rm fm}^{-1}\,,&\quad \gamma=3\, {\rm fm}^{-1}\,,&
     \quad \eta=1.6\, {\rm fm}^{-2}\,.
\end{array}
$$
The RSC potential has a  strong repulsion at small distances,
\begin{eqnarray}
\nonumber
	\|a_{\ell\ell'}\|&=&\frac{2m}{\alpha}\left(
	\begin{array}{cc}
	h_0+h_1+h_2+h_3 &\ 2\sqrt{2}(h_4+h_5+23.5h_0)\\
	2\sqrt{2}(h_4+h_5+23.5h_0) &\ h_1+h_2+h_3-2h_4-2h_5-46h_0-3h_6-3h_7
	\end{array}
	\right)\\
\nonumber \\
\nonumber
 	&\approx&
	\frac{2m}{\alpha}\left(
	\begin{array}{rc}
	6832\, {\rm MeV} &\ -4434\, {\rm MeV}\\
	-4434\, {\rm MeV} &\ 15979\ {\rm MeV}
	\end{array}
	\right)\,.
\end{eqnarray}
In contrast the Moscow potential has very strong attraction instead 
and sustains, apart from the deuteron bound state, a very deep bound state 
known as Pauli Forbidden State (PFS). In this case the expansion 
(\ref{wzero}) begins from the second term,
$$
	\|b_{\ell\ell'}\|= 2m\left(
	\begin{array}{cc}
	\displaystyle{V_1+V_2\frac{\gamma}{\beta}}&\
	\displaystyle{6\sqrt{2}V_2\frac{\gamma^3}{\beta^3}}\\
	\displaystyle{6\sqrt{2}V_2\frac{\gamma^3}{\beta^3}} &\
	\displaystyle{V_1+V_2\frac{\gamma}{\beta}(1-6\frac{\gamma^2}{\beta^2})}
	\end{array}
	\right)
	\approx 2m\left(
		\begin{array}{rr}
	-513\, {\rm MeV} &\ -7156\, {\rm MeV}\\
	-7156\, {\rm MeV} &\ 4547\, {\rm MeV}
	\end{array}
	\right)\,.
$$
Both potentials describe the deuteron properties and the
$np$--scattering quite well despite their completely different
short range behaviour. \\

To begin with we consider real energies corresponding to bound and scattering
states. We integrated Eqs. (\ref{ABeq}) by the Runge--Kutta method from
$r_{min}=10^{-4}$\,fm to $r_{int}=1$\,fm with the boundary conditions
$\|A^{(4)}(k,r_{min})\|$ and $\|B^{(4)}(k,r_{min})\|$. Then from  
$r_{int}=1$\,fm we integrated Eqs. (\ref{fpmeq}) up to
$r_{max}=20$\,fm where the functions $F^{(\pm)}(k,r)$ attain their limits
(\ref{jplus}). Repeating such calculations with different
values of the momentum $k$ corresponding to points on the positive imaginary
axis, we found that the equation
\begin{equation}
	\label{Fm0} \det\|F^{(-)}(k,r_{max})\|=0
\end{equation}
is fulfilled at the points $k_0$ given in Table I. The binding energies
and the percentages of the $D$--waves ($D\%$) are also given in this table. 
For comparison Table I contains the energies and $D\%$ obtained
originally in Refs.\cite{RSC,moscow,kukmod} by the authors that constructed
these potentials. The other observables such as the mean square radius and the
electric quadrupole moment of the deuteron are also  the same as given in
Refs.\cite{RSC,moscow}. Due to the presence of the deep unphysical PFS
state in the Moscow potential the deuteron state is an excited one 
and therefore its wave function has  a node at $r_c\sim 0.59$\,fm.\\

For real positive $k$, i.e., for scattering states, we performed calculations 
with the same $r_{min}$, $r_{int}$, and $r_{max}$.
According to  Eq. (\ref{smatr}) the product
$$
	\|F^{(+)}(k,r_{max})\|\cdot\|F^{(-)}(k,r_{max})\|^{-1}=\|S(k)\|
$$
gives us the $S$--matrix which contains information about the
scattering observables. The nucleon--nucleon $S$--matrix usually is
parametrised in terms of the so--called {\it bar} phase shifts and mixing
parameter, introduced in Ref. \cite{bar}, as follows
$$
	\left(
	\begin{array}{ccc}
	S_{00} & \phantom{,} & S_{02} \\
	S_{20} & \phantom{,} & S_{22}
	\end{array}
	\right)=\left(
	\begin{array}{ccc}
	\displaystyle{{\rm e}^{2i\bar\delta_0}\cos 2\bar\varepsilon} &
	\phantom{,} &
	\displaystyle{i{\rm e}^{i(\bar\delta_0+\bar\delta_2)}\sin 2
	\bar\varepsilon} \\
	\displaystyle{i{\rm e}^{i(\bar\delta_0+\bar\delta_2)}\sin 2
	\bar\varepsilon} &\phantom{,} &
	\displaystyle{{\rm e}^{2i\bar\delta_2}\cos 2\bar\varepsilon}
	\end{array}
	\right)\,.
$$
>From this matrix equation one gets
\begin{eqnarray}
\nonumber
	\bar\delta_0 &=& \frac{\ln S_{00}}{2i}\sqrt{1-
	\frac{S_{02}^2}{S_{00}S_{22}}}\,,\\
\nonumber
	\bar\delta_2 &=& \frac{\ln S_{22}}{2i}\sqrt{1-
	\frac{S_{02}^2}{S_{00}S_{22}}}\,,\\
\nonumber
	\bar\varepsilon &=& \frac14\arccos\left(
	\frac{S_{00}S_{22}+S_{02}^2}{S_{00}S_{22}-S_{02}^2}\right) \,.
\end{eqnarray}
The obtained phase shifts and  mixing parameters are in agreement with
the values given in Ref.\cite{RSC}, for all collision energies
examined (up to $E_{c.m.}=176$\,MeV).
For example, at $E_{c.m.}=12$\,MeV ($k=0.53793\,{\rm fm}^{-1}$) the RSC
potential gives\footnote{The calculated $\|F^{(+)}(k,r_{max})\|$
and $\|F^{(-)}(k,r_{max})\|$ are complex conjugate to each other at least
within five digits}
$$
	\|F^{(\pm)}(k,r_{max})\|= \left(
		\begin{array}{rr}
	0.16247\times10^{11} \mp i0.18386\times10^{12} &\
	-0.60905\times10^{5}\pm i0.68923\times10^{6}\\
	-0.13390\times10^{13}\pm i0.61083\times10^{11} &\
	0.50194\times10^{7}\mp i0.22898\times10^{6}
		\end{array}
	\right)
$$
and
$$
	\|S(k)\|=\left(
		\begin{array}{rcr}
	-0.95640+i0.28507 &\ -0.06232+i0.01229\\
	-0.06232+i0.01229 &\  0.99303-i0.09933
		\end{array}
	\right)\,.
$$
>From the latter  $S$--matrix we obtain the  following scattering parameters
$$
	\bar\delta_0=1.4288\,,\qquad \bar\delta_2=-0.04995\,,\qquad
	2\bar\varepsilon=0.06357\,,
$$
which practically the same  with those obtained by Reid \cite{RSC}\\
$$
	\bar\delta_0=1.426\,,\qquad \bar\delta_2=-0.050\,,\qquad
	2\bar\varepsilon=0.064\,,
$$
via direct a solution of the Schr\"odinger equation.\\

The Moscow potential at the same energy gives different Jost matrices,
$$
	\|F^{(\pm)}(k,r_{max})\|=\left(
	  \begin{array}{rr}
		-361.16 \pm i4347.4 & \ -0.08458\pm i1.0183\\
		30864 \mp i1423.1 & \ 7.2291 \mp i0.33331
	  \end{array}
	\right)\,,
$$

but practically the same $S$--matrix
$$
	\|S(k)\|=\left(
	  \begin{array}{cc}
		-0.95557+i0.28772 &\ -0.06283+i0.01252\\
		-0.06283+i0.01252 &\ 0.99268-i0.10059
	  \end{array}
	\right)\
$$
and therefore the same phase shifts and mixing parameter
$$
	\bar\delta_0=1.4275\,,\qquad \bar\delta_2=-0.05059\,,\qquad
	2\bar\varepsilon=0.06411\,.
$$
The huge values of the above Jost matrix elements are due to the  
behaviour of the potentials at small distances, which
generates large values of the derivatives $\partial_r\|F^{(\pm)}(k,r)\|$ 
that pushes up the absolute values of the functions 
$F^{(\pm)}_{[\ell][\ell']}(k,r)$. To demonstrate an opposite example, 
we solved Eqs. (\ref{fpmeq}) at the 
same energy ($E_{c.m.}=12$ MeV) with a
rudimentary nucleon--nucleon potential (also in the deuteron channel)
consisting of two Yukawa--type terms \cite{blatt},
\begin{equation}
\label{blattpot}
	V(r)=v_c\frac{\exp(-\omega r/\rho_c)}{r/\rho_c}+
	v_t\frac{\exp(-\omega r/\rho_t)}{r/\rho_t}S_{12}\,,
\end{equation}
\begin{eqnarray}
\nonumber
	&& v_c=-22.7\, {\rm MeV}\,,\qquad v_t=-10.9\, {\rm MeV}\,,\\
\nonumber
	&&  \omega=2.12\,,\qquad \rho_c=2.47\, {\rm fm}\,,\qquad
	\rho_t=3.68\, {\rm fm}\,.
\end{eqnarray}
Like the RSC it is singular  at $r=0$, but the coefficients
$$
	\|a_{\ell\ell'}\|=2m\rho_c\left(
	\begin{array}{rcr}
	-22.7\, {\rm MeV} &\ -45.9\, {\rm MeV}\\
	-45.9\, {\rm MeV} &\  9.8\, {\rm MeV}
	\end{array}
\right)
$$
for its expansion (\ref{wzero}) are in two orders of magnitude less than
those for the RSC potential. As a result the Jost matrices calculated at
$E_{c.m.}=12$\,MeV are
$$
	\|F^{(\pm)}(k,r_{max})\|=\left(
	\begin{array}{rr}
		0.27328\pm i0.87845 &\ 0.11204 \pm i0.23730\\
		3.8437\mp i0.37653  &\ 1.6876\mp i0.15658
	\end{array}
	\right)\,.
$$
while the $S$--matrix is
$$
	\|S(k)\|=\left(
		\begin{array}{rr}
	-0.98755+i0.10546 & \ 0.11564 -i0.015980\\
	0.11564 -i0.015980& \ 0.97914-i0.16631
		\end{array}
	\right)\
$$
are $\bar\delta_0=1.53$, $\bar\delta_2=-0.0847$, and
$\bar\varepsilon_0=0.117$.\\

So far we dealt with  momenta on and above the real axis $({\rm Im\,}k\ge0)$ 
and therefore  the coordinate rotation (\ref{rot}) was not needed. Consider 
now a point $k$ which is under the  real axis. To obtain the Jost matrix 
$\|{\cal F}^{(-)}(k)\|$ in this domain of the $k$--plane, we
must integrate the rotated equations (\ref{fpmrot}) (with $\theta>0$)
along the real variable $x=|r|$. Similarly to the case with $\theta=0$, 
in the immediate vicinity of the point $x=0$ it is convenient,
for  numerical reasons, to replace $\|F^{(\pm)}\|$ by their linear 
combinations (\ref{AB}) and the equations (\ref{fpmrot}) by the corresponding 
linear combinations of them. The resulting rotated equations for $\|A\|$ 
and $\|B\|$ as well as the rotated iterative equations follow immediately 
from (\ref{ABeq}), (\ref{A0B0}), and (\ref{ABiter}) after simple replacement 
of $r$ by $x\exp(i\theta)$.\\

To demonstrate the ability of the method to deal with momenta of the fourth
quadrant of the $k$--plane, we calculated the Jost
matrix for the RSC potential at $k=0.5\exp(-i0.3\pi)\,{\rm fm}^{-1}$, i.e.,  
at $E_{c.m.}\approx(-3.20-i9.86)\,{\rm MeV}$, with three different values of
the rotation angle: $\theta=0,\ 0.35\pi,\ 0.4\pi$. The results obtained
are given in Table II. The first line of this table demonstrates that
the unrotated equations cannot give a correct $\theta$--independent Jost 
matrix when ${\rm Im\,}k<0$. The matrix obtained with $\theta=0$ is 
also $r$--dependent, i.e., has no limit when  $|r|\to\infty$.
If however $\theta$ is large enough, such that
${\rm Im\,}kr\ge0$ and  the point $k$ is above the dividing line, then
$\|F^{(-)}(k,x_{max}{\rm e}^{i\theta})\|$ does not depend on $\theta$
(compare the second and third lines of the table) and $x_{max}$. To 
achieve the $x_{max}$--independence when $\theta\ne 0$, we have to go 
further afield because the potential vanishes along the $x\exp(i\theta)$ 
slower than along the real $r$. Thus, the results given in Table II were 
obtained with $x_{max}=50$\,fm.\\

The number of digits which are unchanged under the rotation show the
accuracy achieved. In this connection it is interesting to note that the
accuracy of the second column of the Jost matrix as well as of the
determinant turn out to be always much higher than that of the  Jost
matrix first column. This is exemplified by the two last lines of Table
II. It is interesting to know that the correct value of the
determinant can be obtained even with crude boundary conditions and large
tolerance of the Runge--Kutta procedure. Thus, the correct binding energy of
the deuteron (given in Table I) can be obtained even with boundary conditions
$\|A^{(1)}(k,r_{min})\|$ and $\|B^{(1)}(k,r_{min})\|$ when the first column
of $\|F^{(-)}(k,r_{max})\|$ is even wrong. The only explanation for this  is
that somehow both elements of this column get the same erroneous term which
cancel out in the determinant. This  observation means that the procedure
of locating  spectral points  is less demanding and  less delicate than 
the calculations of the corresponding wave functions.\\

None of the potentials (\ref{RSCpot},\ref{moscowpot},\ref{blattpot}) 
generates resonances (at least at reasonably low energies).
To the best of our knowledge, non-central potentials generating resonances
have not been published yet.
Thus, in order to demonstrate the ability of our method to  locate Siegert
states we  constructed an artificial potential with  a rich
spectrum. For this we used the well--known central potential
\begin{equation}
\label{cpot}
	V_c(r)=7.5 r^2\exp(-r)\,,
\end{equation}
which is widely used as a testing case for new methods that  locate
resonances (see for example Refs. \cite{nth9607028,maier,mandlsh,yamani}).
It is usually assumed that this potential is given in atomic units. In order 
to be consistent, however, with  the potentials used 
in this work, we assume that it is  given here in MeV. Then for 
$\hbar^2/2m=1/2\,{\rm MeV}\,{\rm fm}^2$,  the numerical values of the
resonance energies are the same in MeV and in atomic units.\\

This potential generates a sequence of $S$--wave resonances
(see Ref. \cite{nth9607028}) which cannot be significantly displaced if we
add very weak interaction in the $D$--wave and a weak $S$--$D$ coupling.
Thus the  non--central potential coupling of the $S$ and $D$ partial 
waves which is represented by the matrix
\begin{equation}
\label{wpot}
	\|W_{\ell\ell'}(r)\|=2m\left(
	\begin{array}{ccc}
	7.5 r^2\exp(-r) & \ -\lambda r^2\exp(-r)\\
	-\lambda r^2\exp(-r) & \  -\lambda r^2\exp(-r)\\
	\end{array}
	\right)\,,
\end{equation}
should generate resonances at least when $\lambda$ is small. 
Bound states can also be generated by increasing $\lambda$
since we have chosen  a negative sign for  the $D$--wave potential
and for the  off-diagonal elements of the matrix 
(\ref{wpot}).\\

Of course the use, as a testing case, of a potential with a weak 
$D$--wave and weak coupling is rather undesirable. Therefore, 
we gradually increased $\lambda$ up
to the value $\lambda=15$\,MeV. The resulting spectrum is given in
Table III and depicted in Fig. \ref{resfig}. It consists of eight bound states
and a sequence of resonances (we show only six of them, nearest to the real
axis). By the large open circles in Fig. \ref{resfig} we also show the
positions of the first three resonances when $\lambda=0$ which coincide
with the resonances of the potential (\ref{cpot}) found in
Ref. \cite{nth9607028}. The small open circles show their movement 
when $\lambda$ is gradually increased in steps of $\Delta\lambda=1$\,MeV.
The digits displayed in Table III are those which do not change when
the rotation angle changes and thus they represent the accuracy of our
calculations.
%
\section{Conclusions}
The present  work is a continuation of a series of papers 
\cite{rakpup,nuovocim,nth9607028} in which a practical method for quantum 
mechanical calculations is developed. The method is based on direct 
calculations of the Jost  function and Jost solutions and is a combination 
of the variable--constant method with the complex coordinate rotation. 
Here we have extended it to include non-central potentials.\\

The proposed method offers a unified way of treating bound, scattering, and
resonance state problems. It is a powerful method which enables us to 
investigate the analytical properties of the Jost function in the complex
momentum plane. This opens up new possibilities in locating
resonances. Within this method the Siegert wave functions can be properly
normalised. Even sub-threshold resonances can be located
which is a difficult task for many other methods.
Moreover the formalism  presented  can be easily extended to
complex values of the angular momentum $\ell$ and therefore
Regge trajectories can also be located \cite{nth9607028}.\\

The wave function can be obtained in a form which guarantees its correct
asymptotic behaviour for all three types of physical problems. In all 
cases  the same accuracy can be achieved which can be 
reliably controlled by simply changing the rotation angle.\\

Despite the restriction (\ref{Vinf}),  the potentials
with Coulomb tails can be incorporated into the proposed method in a
straightforward way as it was done in our previous publications
\cite{nuovocim,nth9607028}. To do this we need only to replace in 
all formulae the Riccati--Bessel functions by the corresponding Coulomb 
functions.\\

The method can be extended further: Firstly, to treat the $N$--body coupled
hyperradial equations which differ from the two--body radial equations only
by the possibility of having half--integer values of $\ell$; secondly, to
investigate the  behaviour of the coupled partial waves when the angular 
momenta is complex valued; thirdly, to treat non-analytical and singular
potentials, and fourthly to treat coupled channel problems with different
thresholds. The work on all these extentions is under way.
%
\section*{Acknowledgements}
Financial support from the  University of South
Africa  and the Joint Institute  for Nuclear Research, Dubna
is greatly appreciated.
\newpage
\appendix

\section{Regular Basis}
 By definition, $\|\Phi(k,r)\|$ is a regular basis (fundamental matrix) if
it has the following three properties \cite{newton}:
\begin{itemize}
\item[i)]
Each column of $\|\Phi_{[\ell'][\ell]}\|$ is a solution of
Eq.~(\ref{regeq}),
\item[ii)]
$\Phi_{[\ell'][\ell]}(k,0)=0,\qquad \forall\ [\ell'],[\ell]$\,,
\item[iii)] all columns of $\|\Phi_{[\ell'][\ell]}\|$ are linearly
    independent.
\end{itemize}
Moreover, $\|\Phi\|$ must be a square matrix since
Eq.~(\ref{regeq}) has as many independent regular column--solutions as the
column dimension \cite{kamke}.
\subsection{Behaviour at short distances}
The property ii) implies that the  boundary condition (\ref{zphi})
should be imposed at the point $r=0$. Eq.~(\ref{regeq}), however,
is singular at the origin, and consequently the existence and 
uniqueness theorem\cite{kamke} is not valid at this point. This theorem is 
valid for all  $r \ge \delta > 0$ for any arbitrary  small $\delta$.
Hence, the matrix $\|\Phi\|$ can be uniquely defined by
$\|\Phi(k,\delta)\|$ and $\partial_r\|\Phi(k,\delta)\|$, while  within the
interval $[0,\delta]$,  $\|\Phi(k,r)\|$ (having the above three
properties) can be obtained explicitly as follows.
After multiplying by $r^2$ and using (\ref{V0}), Eqs.~(\ref{regeq})
 decouple  giving
\begin{equation}
\label{zordereq}
	\left[r^2\partial_r^2-\ell(\ell+1)\right]\Phi_{[\ell][\ell']}
		(k,r)\approx0,    \quad r\in[0,\delta]\,.
\end{equation}
Each of these equations has two independent solutions  behaving as
$\sim r^{\ell+1}$ and $\sim r^{-\ell}$ and of course the trivial (zero) 
solution. Therefore a regular column  consists only of $\sim r^{\ell+1}$ and 
zero elements. Therefore the only way to construct  the regular matrix
$\|\Phi(k,r)\|$ obeying Eq. (\ref{zordereq}) with linearly independent
columns, is to choose its diagonal as follows
\begin{equation}
\label{zorder}
	\|\Phi(k,r)\|\mathop{\sim}\limits_{r\to 0}\left(
		\begin{array}{cccc}
	r^{\ell_1+1} & 0 & \cdots & 0 \\
	0 & r^{\ell_2+1} & \cdots & 0 \\
	\vdots & \vdots & \vdots & \vdots \\
	0 & 0 & \cdots & r^{\ell_N+1}
		\end{array}
	\right)
\end{equation}
which obviously satisfies the above-mentioned three properties. Higher order
corrections to Eq. (\ref{zordereq}) give, of course, non--zero off-diagonal
elements. Each column of the matrix (\ref{zorder}) is a solution
of Eq. (\ref{zorder}) and therefore has, separately,
the short--range behaviour
\begin{equation}
\label{ascol}
	\left(
		\begin{array}{c}
	\Phi_{[\ell_1][\ell_n]}(k,r)\\
	\Phi_{[\ell_2][\ell_n]}(k,r)\\
	\vdots\\
	\Phi_{[\ell_n][\ell_n]}(k,r)\\
	\vdots\\
	\Phi_{[\ell_N][\ell_n]}(k,r)
		\end{array}
	\right)\mathop{\sim}\limits_{r\to 0}\left(
		\begin{array}{c}
	0\\ 0\\ \vdots\\ r^{\ell_n+1}\\ \vdots\\ 0
		\end{array}
	\right)\,.
\end{equation}
Hence  when $r\to0$ all elements of the left hand side column of 
Eq. (\ref{ascol}) with $[\ell]\ne[\ell']$ vanish faster than the one 
with $[\ell]=[\ell']$. In other words, the off-diagonal elements of 
a column are infinitesimal as compared to the diagonal one.
Therefore we can rewrite Eq. (\ref{zorder}) indicating the higher order
corrections as follows
\begin{equation}
\label{forder}
	\|\Phi(k,r)\|\mathop{\sim}\limits_{r\to 0}\left(
		\begin{array}{cccc}
	r^{\ell_1+1} & o (r^{\ell_2+1}) & \cdots & o
	(r^{\ell_N+1})\\
	o(r^{\ell_1+1}) & r^{\ell_2+1} & \cdots & 
	o(r^{\ell_N+1})\\
	\vdots & \vdots & \vdots & \vdots \\
	o(r^{\ell_1+1}) & o(r^{\ell_2+1})& 
	\cdots & r^{\ell_N+1}
		\end{array}
	\right)\,.
\end{equation}
The inclusion of the higher order terms retains the linear independence of
the columns. Indeed,
$$
	\det\|\Phi(k,r)\|=\prod_ir^{\ell_i+1}+
	o\left(\prod_ir^{\ell_i+1}\right)\,,
$$
because the products involving off-diagonal elements always contain at least
one of the factors $o(r^{\ell_1+1})$, $o(r^{\ell_2+1})$, 
$\cdots$, $o(r^{\ell_N+1})$. Hence
$$
\det\|\Phi(k,r)\|\mathop{\longrightarrow}\limits_{r\to0}
\prod_ir^{\ell_i+1}\,
$$
and on the interval $(0,\delta]$ we can always find at least one point
$r_0$ where $\det\|\Phi(k,r_0)\|\ne0$. This means that the columns of
the matrix (\ref{forder}) are linearly independent on the whole interval
$(0,\delta]$.\\

The structure of the matrix (\ref{forder}) implies that the behaviour of the
regular basis in the immediate vicinity of the point $r=0$ is such that
$$
     \lim_{r\to 0}\displaystyle{\frac{\Phi_{[\ell][\ell']}(k,r)}
     {r^{\ell'+1}}}=\delta_{[\ell][\ell']}\,.
$$
Since we can choose the normalisation constant in each element of the matrix
$\|\Phi(k,r)\|$ independently, we may replace $r^{\ell'+1}$ in the last
equation by $(kr)^{\ell'+1}$ to get the boundary condition in the form
\begin{equation}
\label{fdjd}
 	    \lim_{r\to 0}\displaystyle{\frac{\Phi_{[\ell][\ell']}(k,r)}
     	    {j_{\ell'}(kr)}}=\delta_{[\ell][\ell']}\,.
\end{equation}
To obtain the higher order terms of (\ref{forder}) in an explicit form
we replace $\|\Phi(k,r)\|$ by the combination of the two new unknown matrix
functions $\|F^{(\pm)}(k,r)\|$, as is given by Eq. (\ref{ansatz}), which are
subjected  to the constraint (\ref{lagrange}) and obey Eqs.
(\ref{fpmeq}). Since by definition the matrix $\|\Phi(k,r)\|$ is regular
at $r=0$, we have
$$
	h^{(+)}_{\ell}(kr)F^{(+)}_{[\ell][\ell']}(k,r)
	\ \mathop{\longrightarrow}\limits_{r\to0}
	\ -h^{(-)}_{\ell}(kr)F^{(-)}_{[\ell][\ell']}(k,r)\,.
$$
On the other hand
$$
	\frac{h^{(+)}_{\ell}(kr)}{h^{(-)}_{\ell}(kr)}\equiv
	\frac{j_\ell(kr)+in_\ell(kr)}{j_\ell(kr)-in_\ell(kr)}
	\ \mathop{\longrightarrow}\limits_{r\to0}\ -1\,,
$$
from which we conclude that in the immediate vicinity of the point $r=0$
the matrices $\|F^{(+)}(k,r)\|$ and $\|F^{(-)}(k,r)\|$ become identical,
i. e.,
$$
	\|F^{(\pm)}(k,r)\|\ \mathop{\longrightarrow}\limits_{r\to0}
	\ \|A(k,r)\|\,,
$$
where the matrix $\|A(k,r)\|$ describes their common short--range asymptotic
behaviour. Thus, we obtain 
\begin{equation}
\label{eeeee}
	\Phi_{[\ell][\ell']}(k,r)\ \mathop{\longrightarrow}\limits_{r\to0}
	\ \frac12\left[h^{(+)}_{\ell}(kr)+h^{(-)}_{\ell}(kr)\right]
	A_{[\ell][\ell']}(k,r)=j_\ell(kr)A_{[\ell][\ell']}(k,r)\,.
\end{equation}
Comparing the last two equations with Eq. (\ref{fdjd}), we get the
following boundary conditions for the matrices $\|F^{(\pm)}(k,r)\|$
\begin{equation}
\label{jfpmj}
     	\lim_{r\to 0}\displaystyle{\frac{
	j_\ell(kr)F^{(\pm)}_{[\ell][\ell']}(k,r)}
     	{j_{\ell'}(kr)}}=\delta_{[\ell][\ell']}\,.
\end{equation}
Note that unlike (\ref{fdjd}), these conditions do not demand that 
all elements of the matrices $\|F^{(\pm)}\|$ be regular. Indeed, when 
$\ell > \ell'$  Eq. (\ref{jfpmj}) holds even if
$F_{[\ell][\ell']}^{(\pm)}(k,r)\sim r^{-(\ell-\ell'-\epsilon)}$ with any
$\epsilon > 0$. Hence, the left bottom corners of  the matrices
$\|F^{(\pm)}\|$ may, in principle, have diverging elements.
As can be seen from the structure of the regular basis, Eq. (\ref{ansatz}), 
the functions $F_{[\ell][\ell']}^{(\pm)}(k,r)$ are closely related to the
Jost solutions and thus their singular behaviour is not surprising. The
$\Phi_{[\ell][\ell']}(k,r)$ itself remains always regular due to the
presence of $j_\ell(kr)$ in (\ref{eeeee}) which compensates the diverging
terms.\\

Eq. (\ref{jfpmj}) gives us the boundary conditions in explicit form only for
the diagonal elements of the matrices $\|F^{(\pm)}\|$  while for the
off-diagonal ones it only implies that
$F_{[\ell][\ell']}^{(\pm)}\sim o(j_\ell/j_{\ell'})$. To obtain them
explicitly, let us take indefinite integrals in both sides of Eqs. 
(\ref{fpmeq}). This gives 
\begin{equation}
\label{intc}
	F_{[\ell][\ell']}^{(\pm)}(k,r)={\rm const}\pm\frac{1}{ik}
	\int h_\ell^{(\mp)}(kr)\sum_{[\ell'']}W_{[\ell][\ell'']}(r)
    	\Phi_{[\ell''][\ell']}(k,r)\,dr\,.
\end{equation}
The integration constants  are fixed for the diagonal and the right--top
corner elements by the conditions (\ref{jfpmj}). Indeed, due  to (\ref{V0})
the integrals in (\ref{intc}) for $\ell\le\ell'$ give functions
($\int r^{-\ell}r^{-2+\varepsilon}r^{\ell'+1}\,dr,\quad \varepsilon>0$)
which vanish at $r=0$. Hence to fulfil  (\ref{jfpmj}) we must set
const=1 for $\ell=\ell'$ and const=0 for $\ell<\ell'$. For the
left--bottom elements we still have to  choose the normalisation which 
is fixed by letting ${\rm const}=\delta_{[\ell][\ell']}$. Thus finally
\begin{equation}
\label{intd}
	F_{[\ell][\ell']}^{(\pm)}(k,r)=\delta_{[\ell][\ell']}\pm\frac{1}{ik}
	\int h_\ell^{(\mp)}(kr)\sum_{[\ell'']}W_{[\ell][\ell'']}(r)
    	\Phi_{[\ell''][\ell']}(k,r)\,dr\,,
\end{equation}
where the indefinite integrals should be understood as  primitive functions
since all arbitrary constants have been fixed already.\\

Consider now Eqs. (\ref{intd}) on a small interval $r\in(0,\delta]$,
where all functions under the integral can be replaced by  their
power--series expansions. While this is possible for $h_{\ell}^{(\pm)}(kr)$ and
$W_{[\ell][\ell']}(r)$, for $\Phi_{[\ell][\ell']}(k,r)$ only the diagonal 
elements are known,
$$
	\Phi_{[\ell][\ell]}(k,r)\approx j_\ell(kr)\,,\qquad  r\in(0,\delta]\,.
$$
As can be seen from Eq. (\ref{forder}) the
elements of the matrix $\|\Phi\|$ vanish (when $r\rightarrow 0$)  with
different speeds. In particular, in each row this speed increases from left
to right. At the same time, within each column the element situated on the
matrix diagonal has the lowest vanishing speed. This
means  that the leading term of the series expansion of a column
is a column which is filled with zeros except the diagonal element.
Looking at either  the differential equations (\ref{fpmeq}) or
the integral equations (\ref{intd}), we see that in fact they are independent
equations for each column. The series expansion of $\|\Phi\|$ should therefore be
individually constructed  for each column. Thus the leading term
$\|\Phi^{(0)}\|$ of such an expansion is
\begin{equation}
\label{Phi0}
	\Phi_{[\ell][\ell']}^{(0)}(k,r)= j_\ell(kr)\delta_{[\ell][\ell']}\,,
	\qquad  r\in[0,\delta]\,,
\end{equation}
which, according to (\ref{eeeee}), implies that 
\begin{equation}
\label{F0}
	F_{[\ell][\ell']}^{(\pm)(0)}(k,r)= \delta_{[\ell][\ell']}\,,
	\qquad  r\in[0,\delta]\,.
\end{equation}
Substituting (\ref{Phi0}) into the indefinite integral (\ref{intd}) and
using the series expansions of $h_\ell$, $j_\ell$, and $W_{[\ell][\ell']}$, we
obtain the first iteration $\|F^{(\pm)(1)}\|$ for all elements of the matrices
$\|F^{(\pm)}\|$ and thus we can get $\|\Phi^{(1)}\|$ which
includes the next terms of the expansion of the regular solution. Using this
iterative procedure we can find as many terms of the expansion as needed
by the following recurrence formulae
\begin{equation}
\label{curr1}
	F_{[\ell][\ell']}^{(\pm)(n+1)}(k,r)=\delta_{[\ell][\ell']}\pm\frac{1}{ik}
	\int h_\ell^{(\mp)}(kr)\sum_{[\ell'']}W_{[\ell][\ell'']}(r)
    	\Phi_{[\ell''][\ell']}^{(n)}(k,r)\,dr\,,
\end{equation}
\begin{equation}
\label{curr2}
	\Phi_{[\ell][\ell']}^{(n)}(k,r)=\frac12\left[
	h_{\ell}^{(+)}(kr)F_{[\ell][\ell']}^{(+)(n)}(k,r)+
	h_{\ell}^{(-)}(kr)F_{[\ell][\ell']}^{(-)(n)}(k,r)\right]\,.
\end{equation}
%
%
\subsection{Behaviour at large distances}
To analyse the long--range behaviour of the matrix--functions
$\|F^{(\pm)}(k,r)\|$ we rewrite the system (\ref{fpmeq})
as follows
\begin{equation}
\label{fpmeqf}
	\partial_rF_{[\ell][\ell']}^{(\pm)}(k,r)=\pm \displaystyle{
       	\phantom{+}\frac{1}{ik}h_\ell^{(\mp)}(kr)
       	\sum_{[\ell'']}}W_{[\ell][\ell'']}(r)\Phi_{[\ell''][\ell']}(k,r)
\end{equation}
It is clear that if the limits $\lim\limits_{r\to\infty}\|F^{(\pm)}(k,r)\|$
exist, then the corresponding derivatives $\partial_r\|F^{(\pm)}(k,r)\|$ are
zero at large distances. The converse is valid, however, only if the
derivatives vanish faster than $r^{-1}$. Indeed, when the derivative of a 
function $\varphi(r)$ behaves as
$$
	\frac{{\rm d}\varphi(r)}{{\rm d}r}\mathop{\sim}\limits_{r\to\infty}
	r^{-(\varepsilon+1)}\,,
$$
the asymptotic behaviour  of the function itself can be written as
$$
	\varphi(r)\mathop{\sim}\limits_{r\to\infty}
	\int r^{-(\varepsilon+1)}{\rm d}r={\rm const} + \left\{
		\begin{array}{ll}
	\ln{r}\, & \varepsilon=0\\
	\displaystyle-\frac{1}{\epsilon r^\varepsilon}\,
	& \varepsilon\ne 0\\
		\end{array}
	\right.\,,
$$
which obviously has a finite limit only if $\varepsilon>0$.\\

Therefore, to establish the domains of the $k$--plane where the limits
$\lim\limits_{r\to\infty}\|F^{(\pm)}(k,r)\|$ exist, it is sufficient to 
find out at which areas of this plane the right--hand sides of Eqs. 
(\ref{fpmeqf}) vanish at large distances faster than the Coulomb potential.
For this we, first of all, need to know how the basic solutions
$\Phi_{[\ell][\ell']}(k,r)$ behave in different domains of the $k$--plane
when $r\to\infty$. To this end we consider Eqs. (\ref{regeq}) at large $r$.
If the potential is of short--range (decaying exponentially or faster) these
equations are reduced to the un-coupled Riccati--Bessel equations, viz.,
\begin{equation}
\label{RBeq}
	\left[\partial^2_r+k^2-\ell(\ell+1)/r^2\right]
	\Phi_{[\ell][\ell']}(k,r)\mathop{\longrightarrow}
		\limits_{r\to\infty} 0\,.
\end{equation}
Considering them as one matrix equation, then there are two linearly
independent solutions which can be chosen as
\begin{equation}
\label{indep}
	\|h^{(\pm)}(kr)\|\equiv\left(
		\begin{array}{cccc}
	h^{(\pm)}_{\ell_1}(kr) & 0 & \cdots & 0 \\
	0 & h^{(\pm)}_{\ell_2}(kr) & \cdots & 0 \\
	\vdots & \vdots & \vdots & \vdots \\
	0 & 0 & \cdots & h^{(\pm)}_{\ell_N}(kr)
		\end{array}
	\right)\,.
\end{equation}
Any other solution is a linear combination of them. Therefore
\begin{equation}
\label{regass}
	\|\Phi(k,r)\|\mathop{\longrightarrow}\limits_{r\to\infty}
	\|h^{(+)}(kr)\|\cdot\|C_1\|+\|h^{(-)}(kr)\|\cdot\|C_2\|\,,
\end{equation}
and according to (\ref{richaninf})
\begin{equation}
\label{regasse}
	\|\Phi(k,r)\|\mathop{\longrightarrow}\limits_{r\to\infty}
	\exp(ikr)\|\tilde C_1\|+\exp(-ikr)\|\tilde C_2\|\,.
\end{equation}
If   ${\rm Im\,}kr=0$ then both terms of the last asymptotic relation
are oscillating and therefore the regular solution 
is bounded, i.e.,
\begin{equation}
\label{imzer}
	\left|\Phi_{[\ell][\ell']}(k,r)\right|\le {\rm const},
	\quad \forall\ \ \ell,\ell'\ \quad {\rm when}\ \ r\to\infty\,.
\end{equation}
For ${\rm Im\,}kr\ne 0$, the regular solution diverges due  either to the
first or the second term of  (\ref{regasse}). In this case we
can keep only the diverging term of it, viz.,
\begin{eqnarray}
\label{imb}
	&& \|\Phi(k,r)\|\mathop{\sim}\limits_{r\to\infty}
	\exp(-ikr)\|\tilde C_2\| \qquad {\rm Im\,}kr>0\,,\\
\label{imm}
	&& \|\Phi(k,r)\|\mathop{\sim}\limits_{r\to\infty}
	\exp(+ikr)\|\tilde C_1\| \qquad {\rm Im\,}kr<0 \,.
\end{eqnarray}
There are, however, special points $k_0$ on the $k$--plane, called spectral
points, where by definition the regular solution contains only the first 
term of (\ref{regass}), i.e.,
\begin{equation}
\label{spoint}
	\|\Phi(k_0,r)\|\mathop{\longrightarrow}\limits_{r\to\infty}^{\rm def}
	\exp(+ik_0r)\|\tilde C_1\|\,.\phantom{-}
\end{equation}
Spectral points situated on the positive imaginary axis correspond to bound
states, and those situated under the real axis to resonance states.\\

Though we have  derived Eqs. (\ref{imzer},\ref{imb},\ref{imm}) describing  the
asymptotic behaviour of the regular basis, under the assumption that the
potential is of the short range, they should be valid  also for 
potentials decaying faster than $\sim 1/r$. The general proof, however,
requires more sophisticated mathematical methods which are beyond the
scope of this article.\\

Substituting Eqs. (\ref{imzer}), (\ref{imb}), (\ref{imm}),
and (\ref{spoint}) into (\ref{fpmeqf}) and 
using the asymptotic form  of the Riccati--Hankel
functions, Eq. (\ref{richaninf}), together with the constraint on the
long--range behaviour of the potential, Eq. (\ref{Vinf}), we find that in
general the derivatives of $\|F^{(+)}(k,r)\|$ and $\|F^{(-)}(k,r)\|$ vanish
rapidly enough in the different domains of the $k$--plane. The only points 
where both limits $\lim\limits_{r\to\infty}\|F^{(\pm)}(k,r)\|$ exist
simultaneously are those with ${\rm Im\,}kr=0$ and the spectral points with
${\rm Im\,}kr>0$.\\

For real $r$ then $\lim\limits_{r\to\infty}\|F^{(+)}(k,r)\|$ exists for 
${\rm Im\,}k\le 0$ and at the spectral points which correspond
to bound states, while $\lim\limits_{r\to\infty}\|F^{(-)}(k,r)\|$ 
exists for ${\rm Im\,}k\ge 0$. In this case the dividing line 
that separates the two domains of the $k$--plane
coincides with the real axis. This line can be turned downwards, to 
expose the resonance spectral points, by rotating $r$ as given 
by Eq. (\ref{rot}). Indeed, if $\phi$ is the
polar angle parametrising a point on the $k$--plane
$$
		k=|k|e^{i\phi}\, ,
$$
then by choosing large enough $\theta>0$ we can make ${\rm Im\,}kr$,
\begin{equation}
\label{imkr}
	{\rm Im\,}kr={\rm Im\,}\left(|k|xe^{i(\phi+\theta)}\right)=
	|k|x\sin(\theta+\phi)\, ,
\end{equation}
positive even when $\phi$ is negative, $-\theta\le \phi\le \pi-\theta$,
i.e., when the point $k$ is in the
fourth quadrant. From the last equation is clear that when
$\theta \ne 0$ the dividing line is
	$\left[-\infty e^{-i\theta},+\infty e^{-i\theta}\right]$ 
(see Fig. \ref{rotfig}).\\

It is worthwhile to mention that the border separating the two
domains of the complex $k$--plane is a line only in the case of long--range
potentials obeying the condition (\ref{Vinf}). If however the potential
decays at large $r$ exponentially, then
$\lim\limits_{r\to\infty}\|F^{(+)}(k,r)\|$ exists also within a band above
the dividing line while
$\lim\limits_{r\to\infty}\|F^{(-)}(k,r)\|$ in the symmetric band below this
line. The faster the potential decays the wider this band is.
For example, if the potential decays as $\sim\exp(-\mu r)$ then the right
hand side of the second equation of the system (\ref{fpmeqf}) behaves,
 below the dividing line  $({\rm Im\,}kr<0)$, as
$$
	\partial_r F^{(-)}(k,r)
	\sim e^{ikr}e^{-\mu r}e^{ikr}=e^{i(2{\rm Re\,}kr-\mu{\rm Im\,}r)}
	\exp\left({-2{\rm Im\,}kr-\mu {\rm Re\,}r}\right)\, .
$$
The last exponential function decays at large distances if
$\displaystyle{{\rm Im\,}kr>-\frac{\mu}{2}{\rm Re\,}r}$, i.e., when
\begin{equation}
\label{band}
	|k|\sin(\theta+\phi)>-\frac{\mu}{2}\cos \theta\ .
\end{equation}
If $\theta=0$, this condition reads ${\rm Im\,}k>-\mu/2$ which enables us to
locate also virtual states (spectral points on the negative imaginary axis)
situated not far from the origin.


\begin{table}
\caption{Bound states generated by the RSC and Moscow potentials}
\begin{tabular}{|c||c|c|c||c|c|}
\hline
 &\multicolumn{3}{c||}{our method}&
\multicolumn{2}{c|}{Refs.\cite{RSC,moscow,kukmod}}\\
\hline
potential & $k_0\ ({\rm fm}^{-1})$ & $E_0\ ({\rm MeV})$ & $D\%$ &
$E_0\ ({\rm MeV})$ & $D\%$ \\
\hline
RSC & $i0.2316110$ & $-2.22460$ & 6.470 & $-2.22460$ & 6.470 \\
Moscow& $i3.5571773$ & $-524.741$ & 14.36 & $-524.8$ & -- \\
Moscow& $i0.2316000$ & $-2.22439$ & 6.588 & $-2.2246$ & 6.778 \\
\hline
\end{tabular}
\end{table}
\begin{table}
\caption{Jost matrix for the RSC potential at $k=0.5\exp(-i0.3\pi)\,
{\rm fm}^{-1}$, calculated with $x_{max}=50\,{\rm fm}$
and three different values of the rotation angle $\theta$.}
\begin{tabular}{|c|c|}
\hline
$\theta$ & $\|F^{(-)}(k,x_{max}e^{i\theta})\|^{\mathstrut}_{\mathstrut}$\\
\hline
 & \\
0 &
$\left(
\begin{array}{cc}
22637458292890+i14763693403731 & \ -22818484+i84501579\\
8536691059869+i6193286881927   & \ -10532586+i32492267
\end{array}
\right)$ \\
 & \\
\hline
 & \\
$\phantom{++}\displaystyle 0.35\pi\phantom{++}$ &
$\left(
\begin{array}{cc}
-325291222087+i639117048997 & \ -2294097-i362305\\
145085673415-i1488370181212 & \ 4729536-i1042674
\end{array}
\right)$ \\
 & \\
\hline
 & \\
$\displaystyle 0.40\pi$ &
$\left(
\begin{array}{cc}
-325288820471+i639116901481 & \ -2294097-i362305\\
145081315229-i1488368045635 & \ 4729536-i1042675
\end{array}
\right)$ \\
 & \\
\hline
\end{tabular}
\end{table}
\begin{table}
\caption{Spectral points of the model potential
(see text) with $\lambda=15$\,MeV.}
\begin{tabular}{|c|c|c|c|}
\hline
$\phantom{+}{\rm Re\,}p_0\ ({\rm fm}^{-1})\phantom{+}$ &
${\rm Im\,}p_0\ ({\rm fm}^{-1})$ &
$E_0\ \ ({\rm MeV})$ &
$\Gamma\ \ ({\rm MeV})^{\mathstrut}_{\mathstrut}$ \\
\hline
0 & 4.5581531714 & $-10.3883801672\phantom{00}$ & 0 \\
\hline
0 & 4.023079863\phantom{0} & $-8.09258579\phantom{000}$ & 0 \\
\hline
0 & 3.471206989\phantom{0} & $-6.02463898\phantom{000}$ & 0 \\
\hline
0 & 2.899849780\phantom{0} & $-4.20456437\phantom{000}$ & 0 \\
\hline
0 & 2.305427868\phantom{0} & $-2.65749883\phantom{000}$ & 0 \\
\hline
0 & 1.681898798\phantom{0} & $-1.41439178\phantom{000}$ & 0 \\
\hline
0 & 1.01619035\phantom{00} & $-0.5163214\phantom{0000}$ & 0 \\
\hline
0 & 0.254097\phantom{0000} & $-0.0322826\phantom{0000}$ & 0 \\
\hline
3.44660892\phantom{00} & $-0.53011439\phantom{000}$ &
5.79904590\phantom{00} & 3.6541940\phantom{000} \\
\hline
4.1388078309 & $-0.1467148600\phantom{0}$ & 8.5541025053 &
1.214449223\phantom{0} \\
\hline
4.46521096\phantom{00} & $-0.686071761\phantom{00}$ &
9.73370724\phantom{00} & 6.12691030\phantom{00} \\
\hline
4.744324\phantom{0000} & $-1.332365\phantom{00000}$ &
10.36671\phantom{000000} & 12.64234\phantom{000000} \\
\hline
4.96356\phantom{00000} & $-1.99719\phantom{000000}$ &
10.3241\phantom{0000000} & 19.8263\phantom{0000000} \\
\hline
5.1410\phantom{000000} & $-2.6634\phantom{0000000}$ &
9.6681\phantom{000000} & 27.385\phantom{00000000} \\
\hline
\end{tabular}
\end{table}
\begin{center}
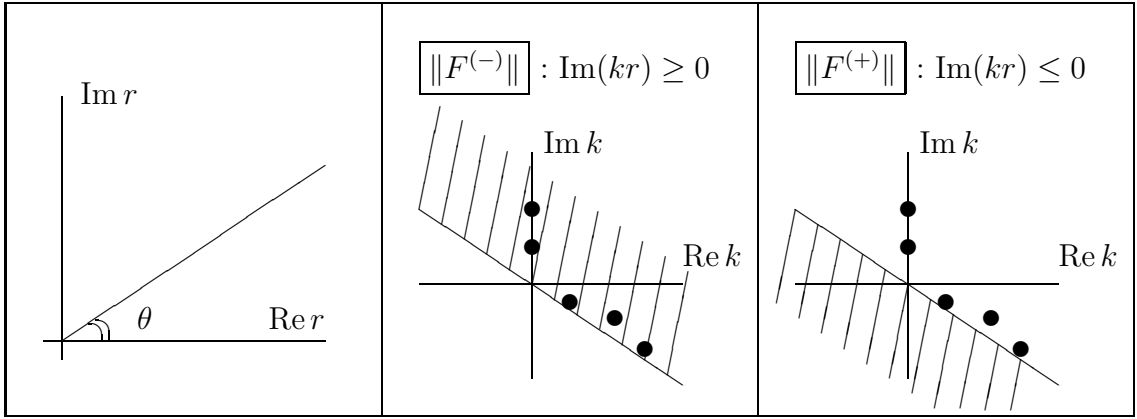
\begin{figure}
\unitlength=0.5mm
\begin{picture}(300,120)
\put(0,10){\line(1,0){300}}    %
\put(0,120){\line(1,0){300}}   %
\put(0,10){\line(0,1){110}}    %
\put(100,10){\line(0,1){110}}  %
\put(200,10){\line(0,1){110}}  %
\put(300,10){\line(0,1){110}}  %
\put(15,30){%
\begin{picture}(0,0)%
\put(-5,0){\line(1,0){75}}
\put(0,-5){\line(0,1){70}}
\put(0,0){\line(3,2){70}}
\put(9,0){\oval(7,11)[tr]}
\put(7,0){\oval(7,9)[tr]}
\put(20,3){$\theta$}
\put(5,63){${\rm Im\,}r$}
\put(70,3){\llap{${\rm Re\,}r$}}
\end{picture}
}
\put(110,100){\fbox{$\|F^{(-)}\|$}\ :\ ${\rm Im}(kr)\ge0$ }
\put(140,45){%
\begin{picture}(0,0)%
\put(-30,0){\line(1,0){70}}
\put(0,-25){\line(0,1){60}}
\put(-30,20){\line(3,-2){70}}
\put(3,35){${\rm Im\,}k$}
\put(40,5){${\rm Re\,}k$}
\put(0,10){\circle*{4}}
\put(0,20){\circle*{4}}
\put(10,-4.8){\circle*{4}}
\put(22,-9){\circle*{4}}
\put(30,-17){\circle*{4}}
\multiput(-30,20)(6,-4){12}{\line(1,5){5.5}}
\end{picture}
}
\put(210,100){\fbox{$\|F^{(+)}\|$}\ :\ ${\rm Im}(kr)\le0$}
\put(240,45){%
\begin{picture}(0,0)%
\put(-30,0){\line(1,0){70}}
\put(0,-25){\line(0,1){60}}
\put(-30,20){\line(3,-2){70}}
\put(3,35){${\rm Im\,}k$}
\put(40,5){${\rm Re\,}k$}
\put(0,10){\circle*{4}}
\put(0,20){\circle*{4}}
\put(10,-4.8){\circle*{4}}
\put(22,-9){\circle*{4}}
\put(30,-17){\circle*{4}}
\multiput(-30,20)(6,-4){8}{\line(-1,-5){5}}
\put(18,-12){\line(-1,-5){4.2}}
\put(24,-16){\line(-1,-5){3.5}}
\put(30,-20){\line(-1,-5){2.7}}
\end{picture}
}
\end{picture}
\caption{Complex rotation of the coordinate and corresponding domains of
the $k$--plane where the limits of the matrices $\|F^{(\pm)}(k,r)\|$
exist.}
\label{rotfig}
\end{figure}
\end{center}
\newpage
\begin{center}
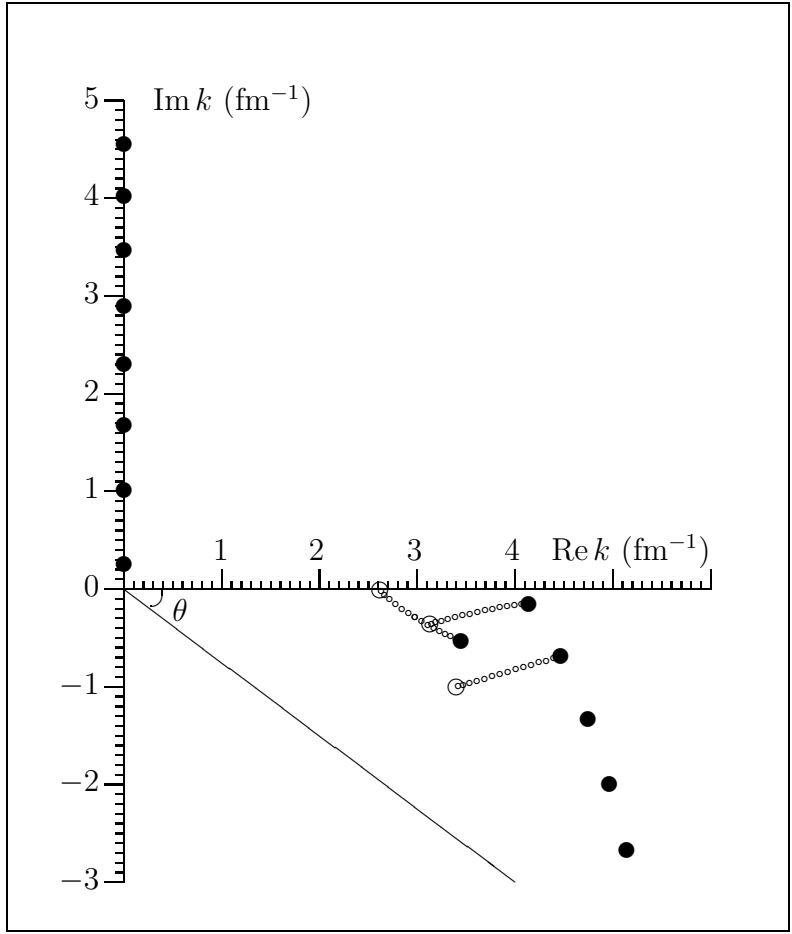
\begin{figure}
\unitlength=1.3mm
\begin{picture}(80,100)
\put(0,5){\line(1,0){80}}         %
\put(0,100){\line(1,0){80}}       %
\put(0,5){\line(0,1){95}}         %
\put(80,5){\line(0,1){95}}        %
\put(12,40){%
\begin{picture}(0,0)%
\put(-2,0){\line(1,0){62}}
\put(0,50){\line(0,-1){80}}
\put(0,0){\line(4,-3){40}}
\multiput(0,50)(0,-10){9}{\line(-1,0){2}}
\multiput(0,49)(0,-1){79}{\line(-1,0){0.8}}
\multiput(10,0)(10,0){6}{\line(0,1){2}}
\multiput(1,0)(1,0){59}{\line(0,1){0.8}}
\put(5.,-3.2){$\theta$}
\put(2.7,0){\oval(2.4,3.5)[br]}
\put(3,49){Im\,$k$ (fm$^{-1}$)}
\put(-2.5,49.3){\llap{5}}
\put(-2.5,39.3){\llap{4}}
\put(-2.5,29.3){\llap{3}}
\put(-2.5,19.3){\llap{2}}
\put(-2.5,9.3){\llap{1}}
\put(-2.5,-0.7){\llap{0}}
\put(-2.5,-10.7){\llap{$-1$}}
\put(-2.5,-20.7){\llap{$-2$}}
\put(-2.5,-30.7){\llap{$-3$}}
\put(60,3){\llap{Re\,$k$ (fm$^{-1}$)}}
\put(10.7,3){\llap{1}}
\put(20.5,3){\llap{2}}
\put(30.5,3){\llap{3}}
\put(40.5,3){\llap{4}}
\put(26.1778617032956307,-0.0487987931002486965){\circle{1.5}}
\put(26.3025462822423650,-0.217686834580884669){\circle{0.5}}
\put(26.6542327975795423,-0.595915758991635985){\circle{0.5}}
\put(27.1695580980224483,-1.05701493791793735){\circle{0.5}}
\put(27.7824888448398122,-1.53852749984262793){\circle{0.5}}
\put(28.4424265688212063,-2.01165723715485439){\circle{0.5}}
\put(29.1159891602262233,-2.46263825984748636){\circle{0.5}}
\put(29.7839609751864742,-2.88471842849552462){\circle{0.5}}
\put(29.7839609751837431,-2.88471842849817028){\circle{0.5}}
\put(30.4369236031901558,-3.27547783546649429){\circle{0.5}}
\put(31.0710605179808574,-3.63552140737563867){\circle{0.5}}
\put(31.6852125981674915,-3.96720590573410914){\circle{0.5}}
\put(32.2793260731891518,-4.27355485781353073){\circle{0.5}}
\put(32.8538001921117528,-4.55759021322850011){\circle{0.5}}
\put(33.4092482730170470,-4.82203308606299230){\circle{0.5}}
\put(33.9464069750364139,-5.06922414945807920){\circle{0.5}}
\put(34.4660892180865330,-5.30114390368393473){\circle*{1.5}}
\put(31.3004244371225226,-3.57144252612435820){\circle{1.5}}
\put(31.4726376361036486,-3.52566398386445379){\circle{0.5}}
\put(31.8979864636724519,-3.41524606157933508){\circle{0.5}}
\put(32.4732972566977418,-3.26406765669983501){\circle{0.5}}
\put(33.1371204669896846,-3.08679132881392870){\circle{0.5}}
\put(33.8551300204522354,-2.89524384182407724){\circle{0.5}}
\put(34.6070269240510209,-2.70007198942068649){\circle{0.5}}
\put(35.3787089000312349,-2.51017000054835904){\circle{0.5}}
\put(36.1590675811612350,-2.33158136109986647){\circle{0.5}}
\put(36.9393913717541267,-2.16730685355656427){\circle{0.5}}
\put(37.7133572970760955,-2.01807768354922423){\circle{0.5}}
\put(38.4767612150451521,-1.88335400367212219){\circle{0.5}}
\put(39.2270282639504009,-1.76206124336912628){\circle{0.5}}
\put(39.9627197481167595,-1.65299994984010767){\circle{0.5}}
\put(40.6831432943956628,-1.55503228615094607){\circle{0.5}}
\put(41.3880783085781001,-1.46714860005315778){\circle*{1.5}}
\put(33.9839242516259032,-9.97251897990203195){\circle{1.5}}
\put(34.1850778528343380,-9.91688356990092834){\circle{0.5}}
\put(34.6764977167069866,-9.78224782693565942){\circle{0.5}}
\put(35.3312017210656437,-9.60136939085920238){\circle{0.5}}
\put(36.0727382915279504,-9.39321861543659486){\circle{0.5}}
\put(36.8578419821196590,-9.16904950620045733){\circle{0.5}}
\put(37.6622135855630580,-8.93579633949382446){\circle{0.5}}
\put(38.4720939462210021,-8.69792617329187667){\circle{0.5}}
\put(39.2796221311301696,-8.45845608144512151){\circle{0.5}}
\put(40.0802859613935691,-8.21951675190577435){\circle{0.5}}
\put(40.8714974065021242,-7.98266548534109055){\circle{0.5}}
\put(41.6517824518253210,-7.74905954148071285){\circle{0.5}}
\put(42.4203165885233524,-7.51955305141412733){\circle{0.5}}
\put(43.1766597550617437,-7.29475455121536664){\circle{0.5}}
\put(43.9206075825836439,-7.07506644951663710){\circle{0.5}}
\put(44.6521096368823933,-6.86071761655368317){\circle*{1.5}}
\put(0,2.54097){\circle*{1.5}}
\put(0,10.1619035){\circle*{1.5}}
\put(0,16.81898798){\circle*{1.5}}
\put(0,23.05427868){\circle*{1.5}}
\put(0,28.99849779){\circle*{1.5}}
\put(0,34.71206989){\circle*{1.5}}
\put(0,40.23079863){\circle*{1.5}}
\put(0,45.581531714){\circle*{1.5}}
\put(47.4432493637802466,-13.3236556657072258){\circle*{1.5}}
\put(49.6356868469191870,-19.9718663611352998){\circle*{1.5}}
\put(51.4023559250308448,-26.6518949871254263){\circle*{1.5}}
\end{picture}%
}
\end{picture}
\caption{Spectral points (filled circles) of the model
potential with $\lambda=15$\,MeV. The large open circles represent
three of the resonances generated by the central potential $V_c$ and
the small circles show the movement of the resonances when 
$\lambda$  decreases from 15\,MeV to zero in steps of 
$\Delta\lambda=1$\,MeV. The dividing line
corresponds to $\theta=0.2\pi$.}
\label{resfig}
\end{figure}
\end{center}


\begin{references}
\bibitem{newton} Newton R G 1967 {\it Scattering Theory of Waves and
		 Particles} (New York: McGraw-Hill)
\bibitem{sasakawa} Sasakawa T and Sawada T 1975 {\it Phys. Rev.} {\bf C 11}
		   87
\bibitem{calog} Calogero F 1967 {\it Variable Phase Approach to Potential
		Scattering} (New York: Academic Press)
\bibitem{rakpup} Pupyshev V V and Rakityansky S A 1994  {\it Z. Phys.}
		 {\bf A 348} 227
\bibitem{ccr} The entire special issue of {\it Int. J. Quantum Chem.}
	      {\bf 14}(4), (1978), is devoted to the complex-coordinate
	      method. See also the review by Ho Y K 1983 {\it Phys. Rep.}
	      {\bf 99} 1
\bibitem{nuovocim} Rakityansky S A, Sofianos S A and Amos K 1996
		   {\it Nuovo Cim.} {\bf 111 B} 363
\bibitem{nth9607028}
	 Sofianos S A and Rakityansky S A 1997
	 {\it J. Phys. A: Math. Gen.} {\bf 30} 3725
\bibitem{fabr} Fabre de la Ripelle M 1983 {\it Ann. Phys.} {\bf 147} 281
\bibitem{gw} Goldberger M and Watson K M 1964 {\it Collision Theory}
		(New-York: Wiley)
\bibitem{palumbo} Palumbo F 1977 {\it Phys.Lett.} {\bf B 69} 275
\bibitem{taylor} Taylor J R 1972 {\it Scattering Theory}
		 (New York: John Wiley \& Sons, Inc.)
\bibitem{abram} Abramowitz M and  Stegun A (ed) 1964
		{\it Handbook of Mathematical Functions}
		(Washington DC: NBS)
\bibitem{mathews} Mathews J and Walker R L 1964 {\it Mathematical Methods of
	      Physics} (New York: W. A. Benjamin, Inc.)
\bibitem{connor} Connor J N L, Smith A D 1983 {\it J. Chem. Phys.}
		 {\bf 78} 6161
\bibitem{landau} Landau L D, Lifshitz E M  1965 {\it Quantum Mechanics}
		 (Oxford: Pergamon)
\bibitem{RSC} Reid R V 1968 {\it Ann. of Phys.} {\bf 50} 411
\bibitem{moscow} Kukulin V I, Krasnopolskii V M, Pomerantsev V N,
		 Sazonov P B 1986 {\it Sov.J.Nucl.Phys.} {\bf 43} 355
\bibitem{kukmod} Michel F and Reidemeister G 1988 {\it Z. Phys. A:
		 Atomic Nuclei} {\bf A 329} 385
\bibitem{bar} Stapp H P, Ypsilantis T J, Metropolis N 1957
	      {\it Phys.Rev.} {\bf 105} 302
\bibitem{blatt} Blatt J M and Weisskopf V F 1952
		{\it Theoretical Nuclear Physics} (New York)
\bibitem{maier} Maier C H, Cederbaum L S and Domcke W 1980
	      {\it J. Phys. B: Atom. Molec. Phys.} {\bf 13} L119
\bibitem{mandlsh} Mandelshtam V A, Ravuri T R and Taylor H S 1993
	      {\it Phys. Rev. Lett.} {\bf 70} 1932
\bibitem{yamani} Yamani H A and Abdelmonem M S 1995
	      {\it J. Phys. A: Math. Gen.} {\bf 28} 2709
\bibitem{kamke} Kamke E 1959 {\it Differentialgleichungen} (Leipzig)
\end{references}
\end{document}